\documentclass[aps,prb,twocolumn]{revtex4-1}
\usepackage{graphicx}
\usepackage{times}
\usepackage{color}
\input pdfcolor.tex
\usepackage{graphicx}
\usepackage{epsfig,color}

\begin{document}

\title{Non-collinear order and gapless superconductivity
in s-wave magnetic superconductors}
\author{Madhuparna Karmakar and Pinaki Majumdar}
\affiliation{Harish-Chandra Research Institute, 
Chhatnag Road, Jhunsi, Allahabad 211 019, India}
 
\begin{abstract}
We study the behavior of magnetic superconductors which involve a local 
attractive interaction between electrons, and a coupling between local 
moments and the electrons.  We solve this `Hubbard-Kondo' model through 
a variational minimization at zero temperature and validate the results 
via a Monte Carlo based on static auxiliary field decomposition of the 
Hubbard interaction. Over a magnetic coupling window that widens with 
increasing attractive interaction the ground state supports simultaneous 
magnetic and superconducting order. The pairing amplitude remains s-wave 
like, without significant spatial modulation, while the magnetic phase 
evolves from a ferromagnet, through non-collinear `spiral' states, to a 
Neel state with increasing density and magnetic coupling. We find that 
at intermediate magnetic coupling the antiferromagnetic-superconducting 
state is gapless, except for the regime of Neel order. We map out the phase 
diagram in terms of density, magnetic coupling and attractive interaction, 
establish the electron dispersion and effective `Fermi surface' in the 
ground state, provide an estimate of the magnetic and superconducting 
temperature scales via Monte Carlo, and compare our results to available 
data on the borocarbides.  
\end{abstract}

\date{\today}
\maketitle

\section{Introduction}

Superconductivity and magnetism are generally competing ordered
states in a material, with superconductivity preferring the 
pairing of time reversed states while magnetism breaks the 
time reversal symmetry. It was argued early on that 
superconductivity and ferromagnetism cannot  coexist 
\cite{ginzburg}.
Externally applied magnetic fields also 
destroy superconductivity - 
either through the generation of a vortex lattice or through the 
Pauli paramagnetic effect \cite{bennemann}.
Magnetic impurities too have a drastic effect
\cite{ag},
with increasing concentration leading quickly to a gapless
superconductor and then the loss of order itself.
These effects seemed to severely restrict the possibility
of superconductivity coexisting with magnetic order.

The situation, however, is more interesting 
and suggestions about 
the {\it coexistence} of superconductivity
and magnetism also date far back.  
In 1963 Baltensperger and Strassler \cite{strassler}
suggested that superconductivity can actually coexist 
with antiferromagnetic order.
Signature of such coexistence was first observed
in the ternary Chevrel phases
\cite{maple_fisher1982,fisher_maple1982, fulde_keller}
RMo$_{6}$S$_{8}$ and RRh$_{4}$B$_{4}$
(where R is a rare earth element). 
In these
materials it is believed that magnetism and superconductivity 
arise from electrons which form distinct subsystems, and the 
ordering of the magnetic degrees of freedom allows the
survival of superconductivity
\cite{lynn2001,lynn1984,thomlinson1982,zwicknagl1981}.

Over the last three  decades many more materials involving 
the interplay of magnetism and superconductivity
have been discovered.
The high $T_c$ cuprates arise from a doped antiferromagnetic
insulator \cite{plee2006}, 
the parent compound of the iron pnictide 
superconductors \cite{stewart_rmp, johnson2010} involves collinear 
antiferromagnetism, the iron chalcogenides \cite{hsu2008, fang2008} 
emerge from a bicollinear antiferromagnetic state, and the 
iron selenides \cite{guo2010, fang2010} also involve proximity 
to an antiferromagnetic insulator.
Over a large part of the phase 
diagram magnetic order coexists with 
superconductivity in these compounds 
\cite{dagotto_natrev,weisenmeyer2011}.
Several heavy fermions also involve 
coexisting magnetic order and superconductivity 
\cite{saxena2000, scalapino_rmp}, {\it e.g},
the Ce compounds \cite{nicklas2007,pham2007}
CeCoIn$_{1-x}$(Cd$_{x}$)$_{5}$ and 
CeIr(In$_{1-x}$Cd$_{x}$)$_{5}$,
and several uranium based
heavy fermions \cite{pfleiderer_rmp}.
In many of these materials electron-electron
repulsion is responsible for emergence of 
local moments and the pairing is usually of
the `off site' $d$-wave type.

A simpler variety of coexistence is seen in the rare earth 
quaternary borocarbides \cite{muller_rev} 
(RTBC), where 
local moments already exist on the rare earths, and Kondo
couple to conduction electrons, and the electrons have a
phonon mediated attraction between themselves.
This is traditional $s$-wave BCS physics playing out
in the background of $f$ moment order, and offers a simple
entry point to the coexistence problem.
Given the similar structure and valence, members of this
family are expected to have the same nominal carrier
density, and electronic structure. What does vary are
the `de Gennes factor' (DG), proportional to ${S(S+1)}$, 
where $S$ is the effective moment on the $f$ ion, and the 
effective pairing interaction, $\eta$, say.  
All materials with a finite DG factor are magnetic but
only compounds with a relatively low DG factor and
larger $\eta$ are superconducting.

Coexisting magnetic and superconducting order 
\cite{maple1995, lynn1997, gupta1998, hilscher_michor1999, 
braun1998, dreschler2009, baba2008, rybaltchencho1999, schultz2011}
have been found in RNi$_{2}$B$_{2}$C 
where, R $=$ Dy, Ho, Er and Tm, in reducing sequence of
the DG factor and increasing $\eta$.
With reducing DG factor the magnetic transition temperature
$T_{AF}$ decreases, from 20K in Gd to 10K in Dy to 2K in Tm,
while the superconducting $T_c$ increases from $\sim$ 6K in
Dy to $\sim$ 11K in Tm. 
$T_{AF}$ scales roughly with the DG factor, and the
magnetic state in all compounds is primarily a
$(0,0,q)$ spiral, while the $T_c$ falls monotonically
with increasing DG factor \cite{muller_rev}. Despite much experimental 
work the detailed symmetry of the paired state, and the
gap anisotropy, is not settled yet.

There is a large theory effort in understanding the 
interplay of magnetism and superconductivity, both in terms 
of general 
phenomenology \cite{blount, greenside, kuper} and specific 
microscopic models \cite{machida1980, grest1981, levin1984, 
nagi1988, umezawa1981, fenton1988, kontani, buzdin1985, 
 buzdin1986, buzdin1981, pines1990, chubukov1998, schamalian2003,
 mckenzie1997, fukuyama1996, schamalian1998, kontani1999, yamada1999, 
 littlewood2001, hertz1976, emery1986, hirsch1987, rink1986, jensen2007}.
Microscopic theories have addressed
the role of magnetic fluctuations in the cuprates 
\cite{pines1990, chubukov1998, schamalian2003},
the layered organics \cite{mckenzie1997, fukuyama1996, schamalian1998,  
kontani1999, yamada1999},
and the heavy fermions \cite{littlewood2001, hertz1976,
emery1986, hirsch1987, rink1986}, to name a few.
We wish to start with the simpler situation, pertinent to
the borocarbides, where one can employ a `Kondo lattice',
for the large $4f$ moments, augmented by a local 
attractive interaction
between the electrons \cite{paiva, pruschke}.

The local moments arising from the $4f$ shell couple to the 
conduction electrons through a Kondo coupling. The ground state 
behavior of such a model has been addressed earlier in one
spatial dimension \cite{paiva} via density matrix renormalisation 
group (DMRG) treating the local moments as $S=1/2$.  There have 
also been studies in higher dimensions \cite{fulde2000, kontani, jensen2007,
jensen2002, maki2004, lee2004, shorikov2006, maki_physica2004} 
aimed at reproducing specific features of the borocarbides
but a general understanding of the interplay of pairing and
magnetic correlations, even in this simple model, appears to be
lacking.

In particular one would have liked to know (i)~how the
magnetic ground state is affected by pairing, (ii)~the 
attraction and Kondo coupling window over which superconductivity
coexists with magnetic order, and (iii)~the spectral features
of the system, given that pairing now occurs between magnetic
Bloch states, and not simply ${\bf k}\uparrow$ and $-{\bf k} 
\downarrow$, and can lead to anisotropic gaps, and even a 
gapless state.

In this paper we report on the ground state of a 
model with s-wave pairing tendency (local attractive
interaction) in the presence of a local moment lattice.
The existence of magnetic moments ${\bf S}_{i}$ is 
predefined, it does not depend on the itinerant electrons
and is independent of the strength of $U$. 

If the  moments are large ($2S \gg  1$) their
quantum fluctuations can be ignored to start with and 
the Kondo effect itself is not relevant. Such a system
can be described by a Kondo lattice of 
`classical' spins coupled to the conduction electrons.
The parameter space of the problem is defined by 
electron density $(n)$, attractive pairing interaction
$(U)$, the `Kondo' coupling $(J)$, and temperature
$(T)$. Most of the results in this paper pertain to
the ground state, the finite 
temperature phase competition will be 
discussed elsewhere.
Our main results are the following
\begin{enumerate}
\item
{\it Magnetic ground state:}
The magnetic ground state depends only weakly on the pairing interaction
and is determined mainly by the electron density and Kondo
coupling, consistent with the suggestions of Anderson and Suhl 
\cite{and-suhl} made originally in the weak coupling context.
\item
{\it Superconducting order:}
At weak Kondo coupling the pairing order parameter increases monotonically 
as $n$ varies from $[0,1]$ but beyond a critical coupling the $n=1$
state loses superconductivity, while it survives for $n \neq 1$ to almost 
twice  the coupling.
\item
{\it Gapless state:} Although the pairing amplitude
is essentially homogeneous, for $n \neq 1$ the superconductor
becomes gapless at a coupling $J_{g}(n,U)$ that is roughly
half of the critical coupling, $J_c(n,U)$, needed for
destroying superconductivity.  At $n=1$ the superconductor 
remains gapped despite the magnetic order.
\item
{\it Quasiparticles and density of states:}
Superconductivity in a generic `spiral' magnetic
background leads to a dispersion with upto eight branches, some 
of which cross the Fermi level for $J > J_g$.
The associated density of states shows multiple van Hove
singularities and the low energy spectral weight maps out
a `Fermi surface' even in the superconducting state.
\item
{\it Comparison with experiments:}
Our ground state is consistent with observations in the borocarbides
and suggest that the superconducting gap in DyNi$_{2}$B$_{2}$C and
HoNi$_{2}$B$_{2}$C could be strongly anisotropic.
\end{enumerate}
The rest of the paper is organized as follows, in 
Section II we discuss our  model and the numerical 
methods.  Section III discusses our results on the phase 
diagram and spectral features obtained within a
restricted variational scheme in two dimensions. 
Section IV compares these results to that from a Monte Carlo
based unrestricted minimization, comments on extensions to
a wider interaction window, and compares our results to
experiments on the borocarbides.

\section{Model and method}

We study the attractive Hubbard model in two dimension
on a square lattice in presence of Kondo like coupling:

\begin{equation}
H = H_{0} - \vert  U \vert  \sum_{i}n_{i\uparrow}n_{i\downarrow}  -
J \sum_{i}{\bf S}_{i}.{\bf \sigma}_{i}
\end{equation}

with, $ H_0 =  \sum_{ij, \sigma}(t_{ij} - \mu \delta_{ij}) 
c_{i\sigma}^{\dagger}c_{j\sigma}$, where $t_{ij}=-t$ for  
nearest neighbor hopping and is zero otherwise. 
${\bf S}_{i}$ is the core spin, arising from $f$ levels,
for example, in a real material. 
${\bf \sigma}_{i}$ is the electron 
spin operator.  $U$ is the attractive onsite interaction
(with a physical origin in local electron-phonon
coupling).
Most of the detailed results in this paper are 
at $U = 4t$, but we have also shown some
results at weaker $U/t$.

This paper focuses on the ground state, which can be
reasonably accessed within mean field theory (MFT), but
we want to set up a scheme that can also access the interplay 
of magnetic and pairing fluctuations at
finite temperature in a situation where $U$ and $J$ are
comparable to $t$. 
While mean field theory can be extended to finite
temperature to access some thermal effects we want
a formulation which (i)~retains the effect
of magnetic fluctuations on pairing, and (ii)~the effect
of the changing low energy electron spectrum on magnetism.
With this in mind we set up a lattice field theory involving 
the electrons and the magnetic and pairing degrees of freedom
as follows.

We apply a single channel 
Hubbard-Stratonovich decomposition to the attractive interaction 
in terms of an auxiliary complex scalar field 
$\Delta_{i}(\tau) = \vert \Delta_{i}(\tau)\vert e^{i\theta_{i}(\tau)}$. 
This converts the `four fermion' term to quadratic fermions in an
arbitrary spacetime fluctuating pairing field.  
On the magnetic side we have a quantum 
`spin S' magnetic moment ${\bf S}_i$ coupled to the 
electrons. 

This problem can be exactly  treated only via
methods like
quantum Monte Carlo. We attempt to retain the thermal
fluctuation effects by (i)~dropping the $\tau$ dependence of $\Delta$ 
but keeping its {\it spatial fluctuations}, and (ii)~treating
${\bf S}_i$ as a classical (large $S$) spin but retaining
its angular fluctuations at finite temperature.
We will discuss the validity of these approximations in
the discussion section.

The pairing field is now `classical', with an amplitude 
$\vert \Delta_i \vert$ and phase $\theta_i$ and the
magnetic moment ${\bf S}_i$ is described in terms of
its polar angle ${\alpha}_{i}$ and azimuthal angle ${\phi}_{i}$.
We set $\vert {\bf S}_i \vert = 1$, absorbing the magnitude of the spin
in the coupling $J$.
The resulting effective Hamiltonian takes the form:
$$
 H_{eff} = H_{0} + 
\sum_{i}(\Delta_{i}c^{\dagger}_{i\uparrow}c^{\dagger}_{i\downarrow}
 + h.c) 
- J\sum_{i}{\bf S}_{i}.{\bf \sigma}_{i} + 
\sum_{i}\frac{\vert  \Delta_{i}\vert^{2}}{U}
$$
where, $\sum_{i}\frac{\vert \Delta_{i}\vert^{2}}{U}$ is the stiffness
associated with the pairing field.
The configurations $\{\Delta_i, {\bf S}_i \}$ 
that need to be considered 
follow the Boltzmann distribution, obtained by tracing over the
electrons:
\begin{equation}
 P\{\Delta_{i}, {\bf S}_i \} \propto Tr_{c,c^{\dagger}}e^{-\beta H_{eff}}
\end{equation}
Physically, the probability of a configuration $\{\Delta_i, {\bf S}_i \}$
is related to the free energy of the electrons in that
configuration. 

To create some insight it is helpful to write down the form of
$P\{\Delta_{i}, {\bf S}_i \}$ expanded to low order in
$\Delta_i $ and $J{\bf S}_i$.
\begin{eqnarray}
P & \propto & 
Tr_{c,c^{\dagger}}
e^{-\beta H_{eff}\{\Delta_{i}, {\bf S}_i \}} \sim 
e^{-\beta {\cal F}_{eff}\{\Delta_{i}, {\bf S}_i \}} \cr
\cr
{\cal F}_{eff} & = & {\cal F}_{\Delta} + {\cal F}_J + 
{\cal F}_{\Delta,J} \cr
\cr
{\cal F}_{\Delta} & = &  \sum_{ij} a_{ij} \Delta_i \Delta^*_j
+ \sum_{ijkl} b_{ijkl} \Delta_i \Delta^*_j \Delta_k \Delta^*_l
+ {\cal O}(\Delta^6) \cr
{\cal F}_J & = & \sum_{ij} J^{(2)}_{ij} {\bf S}_i.{\bf S}_j + \sum_{ijkl}
J^{(4)}_{ijkl} ({\bf S}_i.{\bf S}_j{\bf S}_k.{\bf S}_j + ..) + .. \cr
{\cal F}_{\Delta,J} & = & \sum_{ijkl} [c_{ijkl} \Delta_i \Delta^*_j
{\bf S}_k.{\bf S}_l + h.c] + ..
\nonumber
\end{eqnarray}
where $a_{ij} \sim -\chi^P_{ij} + (1/U)\delta_{ij}$, $\chi^P_{ij}$
being the non-local pairing susceptibility of the free Fermi
system, and $b_{ijkl}$ arises from a convolution of
four free Fermi Green's functions.
$J^{(2)}_{ij} \sim -J^2 \chi^S_{ij}$,
where $\chi^S_{ij}$ is the nonlocal spin susceptibility of the
free electron system, leading to the RKKY interaction, and
$J^{(4)}$, like $b_{ijkl}$, involves a four Fermi cumulant.
$c_{ijkl}$ can be constructed again from a combination of
four Green's functions. \\
The terms above define a relatively
low order classical field theory on a lattice. 
$H_{\Delta}$  involve the first two terms in the
superconducting Ginzburg-Landau 
theory, and $H_J$ describes the leading interaction coupling
magnetic moments. $H_{\Delta,J}$ indicates how the two orders
modify each other.
All of this holds when $\Delta_i$ and $J{\bf S}_i$ 
are $\lesssim t$.

For large and random $\{\Delta_i, J{\bf S}_i \}$ 
the fermion trace can only be evaluated numerically. 
We use two strategies: 
(i)~When considering $T=0$, as in this paper, we can restrict ourselves to
{\it periodic} configurations
of $\{\Delta_i, {\bf S}_i \}$ and in that case we only need to 
estimate the energy of $H_{eff}$ for periodic pairing/magnetic
backgrounds, accomplished readily through the Bogoliubov-de Gennes
(BdG) scheme as we discuss below.
(ii)~When considering finite temperature,
where fluctuations are essential, we generate equilibrium configurations 
by using the Metropolis algorithm for the $\{\Delta_i, {\bf S}_i \}$
and estimate the `update cost' by diagonalizing the 
electron Hamiltonian $H_{eff}$
for every microscopic move.
Needless to say this is a numerically expensive process. 

\subsection{Variational scheme}

As $T \rightarrow 0$ the classical fluctuations die off and 
the fields ${\bf S}_i$ and $\Delta_i$ should be
chosen to minimize the energy.
An unrestricted real space minimization is still a non trivial
task but we choose to minimize the energy using a restricted
family of $\{{\bf S}_i, \Delta_i\}$ configurations, described
below, and check the quality of the result via Monte Carlo
based simulated annealing.
Specifically, we assume 
$\Delta_i = \Delta_0$, a site independent real quantity, and
for the magnetic order we consider 
spiral configurations where the polar angle $\alpha_i = \pi/2$ and
the azimuthal angle $\phi_{i}$ is periodic:
$ S_{zi}  =  0,~
S_{xi} = cos({\bf q}.{\bf r}_i),~
S_{yi}  =  sin({\bf q}.{\bf r}_i) $. 
The allowed wavevectors  $\{ q_x,q_y \}$
are of the form $  2n\pi/L$, where 
($n=1,2,3...$).
We minimize the energy over
$\{ q_x,q_y \} $ and $\Delta_{0}$ for a fixed
$\mu$, $J$ and $U$.

Typically one obtains an unique minimum $\{ \Delta_0, 
{\bf q}\}_{min}(\mu)$.
On this background one calculates the density $n(\mu)$, and then
generates the function $\{ \Delta_0, {\bf q}\}_{min}(n)$. There are
exceptional $\mu$, however, where the minimum is degenerate (for no
symmetry related reason) and one obtains two sets, called
$\{ \Delta_0, {\bf q}\}^+_{min}(\mu)$ and 
$\{ \Delta_0, {\bf q}\}^-_{min}(\mu)$, say. These lead to densities
$n^+(\mu)$ and $n^-(\mu)$, with a discontinuity $\delta n
= n^+ - n^-$. The abrupt change in the background indicates
a first order transition, and the density discontinuity defines
the window of phase separation in the phase diagram.
A constant $n$ minimization would
not have identified it.

A further lowering of energy is possible if
a periodic component is superposed
on $\Delta_0$ but this non-uniform component is
small in the parameter space we explore
\cite{buzdin1985}. Also, in the ferromagnetic window,
where the exchange $J{\bf S}_i$ generates an effective
uniform internal field, a modulated FFLO state can arise.
We quantify this effect separately.

The variational scheme was tested on sizes upto $30 \times 30$
and give stable results for $U \gtrsim 2t$. Although the VC is
doable for larger sizes we did not attempt that since we
wanted comparison with a Monte Carlo based minimization
(see below).

\subsection{Unrestricted minimization}

In addition to the variational scheme we have employed 
the Monte Carlo technique as a simulated annealing tool 
to obtain the ground state, without imposing any
periodicity on the spins or any homogeneity on the
$\Delta_i$.  For this the system is cooled down 
from an uncorrelated high temperature state.
Owing to the computational cost in diagonalizing the 
$ 4L^2 \times 4L^2$ matrix involved in this study most 
of the Monte Carlo 
simulations are done on system size $16 \times 16$, 
and some on $24 \times 24$.

In the discussion section we
compare the ground state phase diagram  obtained through 
our restricted variational scheme with that obtained through 
the `unrestricted' minimization via Monte Carlo.
The 
agreement is reasonable and for the moment we focus on the
variation based phase diagrams.

\subsection{Green's function for the spectrum}

Within the variational scheme the magnetic-superconducting
background has a translational symmetry so the corresponding
electron problem can be diagonalised in momentum and spin
space. For a given ${\bf k}$ the 
BdG problem in the periodic
background involves a $8 \times 8$ matrix and it is
difficult to extract information about the eigenvalues,
and the resulting density of states, analytically. 

However, if $\Delta_0,~J \ll zt$, where the coordination 
number $z=4$ in 2D, one can set up a useful low order
approximation for the Green's function of the electron.
For an electron propagating with momentum ${\bf k}$ 
and spin up, the magnetic scattering connects it to
an electron state with ${\bf k} + {\bf Q}, \downarrow$,
while the pairing field connects it to a hole with
$-{\bf k}, \downarrow$. The matrix elements are,
respectively, $J$ and $\Delta_0$. This leads to the
the Green's function:
\begin{eqnarray}
G_{\uparrow \uparrow}({\bf k}, i\omega_n)
& =&  {1 \over {i \omega_n - (\epsilon({\bf k}) -\mu) 
- \Sigma_{\uparrow \uparrow}({\bf k}, i\omega_n)}} \cr
~~&&\cr
~~&&\cr
\Sigma_{\uparrow \uparrow}({\bf k}, i\omega_n) & = & 
{\Delta_0^2 \over {i \omega_n + (\epsilon({\bf k}) -\mu)}}
+ {J^2 \over {i \omega_n - (\epsilon({\bf k} + {\bf Q}) -\mu)}} 
\nonumber
\end{eqnarray}
where $\epsilon({\bf k}) = - 2t(cos(k_x) + cos(k_y))$.
The self energy of course has higher order terms involving $J^2 \Delta_0^2$,
{\it etc}, but the form above is surprisingly accurate except at $n=1$.
We can extract the spectral function $A_{\uparrow \uparrow}({\bf k}, \omega)
= -(1/\pi) Im~G_{\uparrow \uparrow}({\bf k}, \omega + i \eta)
\vert_{\eta \rightarrow 0} $. A similar expression can be used for
$ A_{\downarrow \downarrow}({\bf k}, \omega) $.
We discuss the comparison of these results with full BdG later on.

\subsection{Computation of observable}

At $T=0$ for a fixed choice of $U$, $J$ and $\mu$ 
the state is characterized by 
the pairing order parameter $\Delta_0$ and the magnetic
wavevector ${\hat x}Q_x + {\hat y} Q_y$. These are
determined by energy minimization. In this periodic 
background we
compute the following:
(i)~the spin and momentum resolved spectral function,
$A_{\sigma \sigma }({\bf k}, \omega)$, from a 
knowledge of the BdG eigenvalues and eigenfunctions,
(ii)~the total electronic density of states $N(\omega)
= \sum_{{\bf k}, \omega} A_{\sigma \sigma }({\bf k}, \omega)$,
(iii)~the overall gap, from the minimum eigenvalue in the
BdG spectrum, (iv)~momentum dependence of the $\omega = 0$
spectral weight, $ \sum_{\sigma}
A_{\sigma \sigma }({\bf k}, 0)$, mapping out the
`Fermi surface' in the superconductor.

While the numerical results for these are based on the 
full BdG numeric, we use the simple Green's function
scheme outlined earlier to explain the physical basis
of the effects.

%-----------------------------------------------------------------
\begin{figure}[b]
 \centerline{
 \includegraphics[height=6cm,width=4.5cm,angle=0]{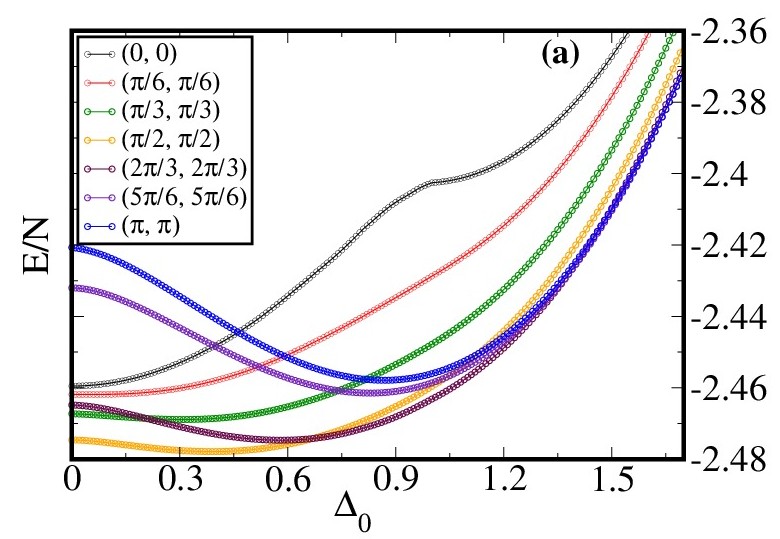}
 \includegraphics[height=6cm,width=4.5cm,angle=0]{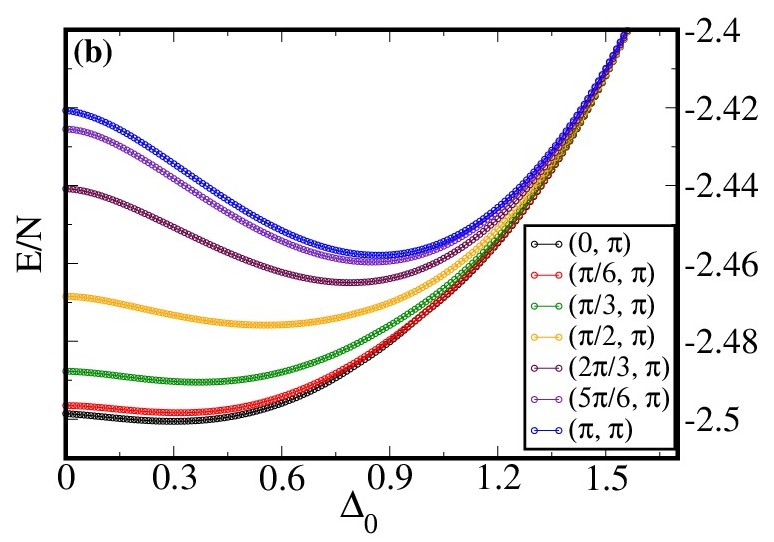}
 }
\caption{Color online: Dependence of the energy on the
pairing field, at $U = 4t$, $J = 1.0t$ and $n \sim 0.4$,
for magnetic wavevectors ${\bf q} = (q_x,q_y)$.
(a)~${\bf q} =  \{q, q \}$ and (b) $\{q, \pi\}$.
The optimized state
is obtained by computing the energy for all possible
${\bf q}$ in the Brillouin zone.
}
\end{figure}
%-----------------------------------------------------------------
\begin{figure}[t]
\centerline{
\includegraphics[height=5.0cm,width=6.5cm,angle=0]{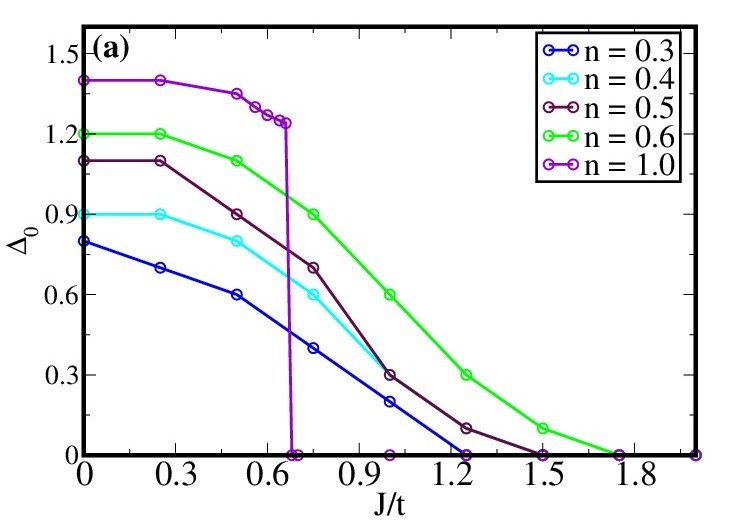} }
\centerline{
\includegraphics[height=5.0cm,width=6.5cm,angle=0]{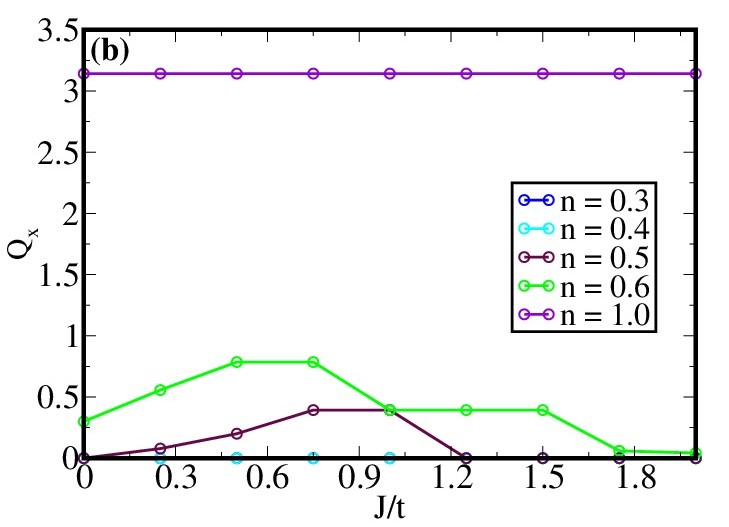} }
\centerline{
\includegraphics[height=5.0cm,width=6.5cm,angle=0]{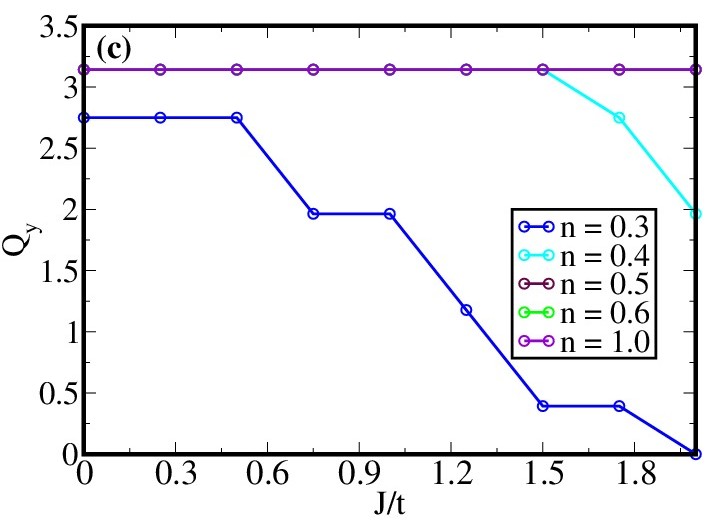} }
\caption{Color online:
Filling dependence of
the optimized
(a)~pairing field amplitude
$\Delta_{0}$ and (b)-(c)~components of the magnetic
wavevector ${\bf Q}$, at different
magnetic interactions $J/t$, and density $n$,
for $U/t = 4$. For $n \neq 1$ the
pairing field undergoes a second order transition with increasing
$J$, while
at $n = 1$ a first order transition is observed.
}
\end{figure}
%-----------------------------------------------------------------

\section{Results}

We organize the results in terms of the thermodynamic phase diagram,
mapping out the magnetic order and superconductivity, and the 
quasiparticle properties which dictate the low energy spectral
features.

\subsection{Phase diagram}

\subsubsection{Energy minimization}

We start with results on the 
dependence of the energy on  $\Delta_{0} $ for
different choices of ${\bf q}$.
At a given $U$ the 
optimized $\Delta_0(\mu,J,U)$ is finite for $J < J_c(\mu,U)$
and falls monotonically as $J$ increases from zero. At weak
$J$ the associated magnetic ordering wavevector ${\bf Q}(\mu,J)$
almost tracks the free band 
RKKY result 
{\it even if $U$ is large}, except near $n=1$.

We will discuss the general features further on 
and for the
moment focus on $E(\Delta_0, {\bf q}) = 
\langle H_{eff}(\Delta_0, {\bf q}) \rangle$ 
at a typical parameter point: $U=4t$, $J=t$ and $\mu = -2t$
(corresponding roughly to $n=0.4$), in Fig.1.
The figure shows the variation of the energy with respect to 
$\Delta_0$ for different choices of ${\bf q}$ (covering panels (a) and (b))
and the absolute minimum defines the appropriate magnetic-superconducting
state. The ground state phase diagram is established by carrying out
this exercise for different $\mu$, $J$ and $U$.

Given our parametrisation of the variational state, we always have
magnetic order with some ${\bf Q}$ (where ${\bf Q}$ denotes the
optimized value of ${\bf q}$), while superconducting order is
absent if the optimum $\Delta_0=0$.

%-----------------------------------------------------------------
\begin{figure*}
 \centerline{
 \includegraphics[height=5.7cm,width=5.6cm,angle=0]{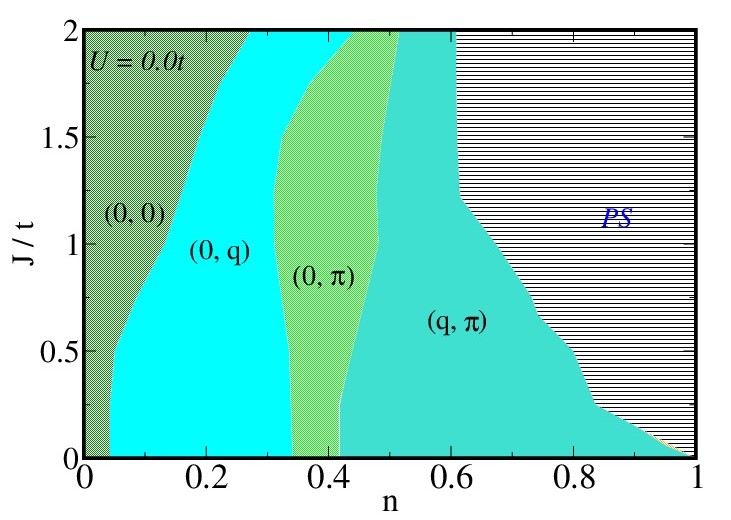}
\hspace{-.1cm}
 \includegraphics[height=5.5cm,width=5.5cm,angle=0]{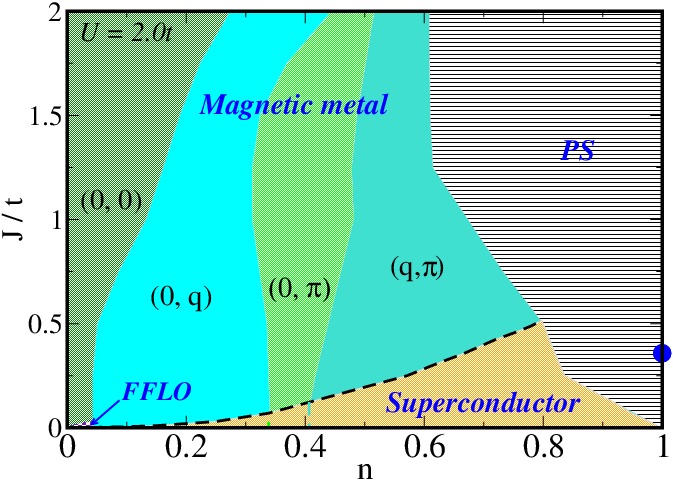}
\hspace{.1cm}
 \includegraphics[height=5.5cm,width=5.5cm,angle=0]{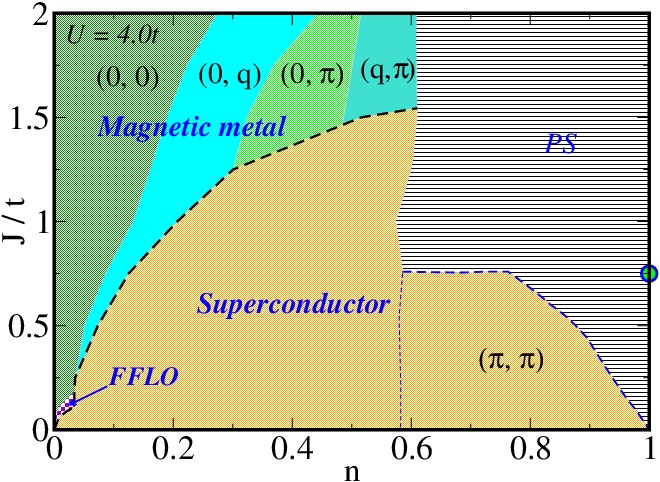}
 }
\caption{Color online:
Ground state $n-J$ phase diagrams showing
evolution of the magnetic and superconducting phases for
three values of $U$.
(a)~The purely magnetic phase diagram at $U=0$.
The magnetic phase changes with the filling but the
order of the occurrence of phases remains unchanged
with varying $J$. (b)~At $U=2t$ superconductivity is
seen over a $J$ window that widens with increasing $n$. The
magnetic phases remain roughly as they were at $U=0$.
(c)~At $U=4t$ the superconducting window is wider, and the
magnetic phases near $n=1$ are modified although elsewhere
it looks roughly similar to the small $U$ picture.
There is a tiny window of modulated superconducting
order (FFLO) state, in the bottom left corner of the finite
$U$ phase diagrams (see text) but they are almost invisible
on the $n-J$ scales used here.
}
\end{figure*}
%-----------------------------------------------------------------

\subsubsection{Variation of pairing field and magnetic order}

Tracking the minimum for varying $\mu$ and $J$ leads to the
ground state parameters shown in Fig.2. 
Over the $J$ range that we explore the density $n(\mu,J)$ (to an accuracy
$\sim 0.01$) is almost independent of $J$ at fixed $\mu$. That allows us to
phrase the results in terms of $n$, although the minimization
was done at fixed $\mu$ and $J$.
Fig.2(a) shows the $J$ dependence of the pairing field amplitude
at several values of $n$. The results here are for $U=4t$, we will
discuss the phase diagram at other values of $U$ later.

Increase in magnetic 
coupling suppresses the pairing field amplitude. 
At a scale $J_c(n)$ 
the pairing amplitude vanishes, indicating 
the destruction of the superconducting phase. 
We make a few observations:
(1)~$J_c(n)$ vanishes as $n \rightarrow 0$, and it   
increases with $n$ with a maximum at $n \sim 0.6$ at
$U=4$. This maximum, $J_c^{max}$, $ \sim 1.5t$.
(2)~The critical
value at $n=1$ is much smaller, with $J_c(n=1) \sim 0.75t$.
(3)~The transition with increasing $J$ is first order at 
$n=1$ and second order for $n \neq 1$.

Fig.2(b) and 2(c) shows the components of the corresponding magnetic
wave vectors. In the absence of pairing, and at low $J$, the
magnetic order is decided by the RKKY interaction, with the 
peak in the band susceptibility $\chi_0({\bf q})$ dictating
the ordering wavevector ${\bf Q}$.
At larger $J$ the spiral states gradually give way to
collinear phases and finally to just two phases,
ferromagnetic and Neel, with a window of phase
separation in between.
In the presence of a pairing interaction it is not
essential that the same trend be followed but, as 
pointed out long back by Anderson and Suhl \cite{and-suhl}, the
presence of pairing affects the electronic density
of states only over a window $2 \Delta_0 \ll \epsilon_F$
so except for ${\bf q} \rightarrow 0$ the spin susceptibility
is mostly unaffected. 

Our results are at $U=4t$ with the  pairing
field $\Delta_0 \sim t$ so the density of states is affected
over a fairly wide window. Nevertheless, except near $n=1$,
the RKKY trend still holds at small $J$. The phase diagrams
in Fig.3 quantify these further.

\subsubsection{$n-J$ phase diagrams}

Fig.3 shows the ground state phase diagram obtained 
through our variational calculations.
The $U = 0$ situation, panel (a),
corresponds to just the 
classical Kondo lattice in two dimensions.
With respect to this non superconducting reference,
(b) and (c) show the impact of increasing pairing
interaction on the magnetic state as well as the
increasing window of superconducting order.
We discuss the three cases separately.

{\bf (i)}~{\it No pairing interaction $(U=0)$:}
In this case $\Delta_0 =0$ and the ground state is
characterized only by ${\bf Q}$. We discuss
the $J/t \rightarrow 0$ and the $J/t \gtrsim 1$ 
limits separately.

The small $J/t$ limit is controlled by
the RKKY interaction with the effective spin-spin
coupling being  $J_{ij} \propto J^2 \chi^0_{ij}$, 
where $\chi^0_{ij}$ is the non local band
susceptibility of the conduction electrons. 
The ordering wavevector is dictated by the
maximum in $\chi^0({\bf q})$, the Fourier
transform of $\chi^0_{ij}$. This depends on $\mu$,
or the filling $n$. 
The system evolves from
a ${\bf Q} = \{0,0\}$ (ferromagnet) at low filling,
to a $\{0,  q\}$ phase at the intermediate filling. 
Further increase in 
filling leads to a $\{0, \pi\}$ antiferromagnet, followed 
by a $\{q, \pi\}$ phase and then to a  $\{\pi, \pi\}$ Neel
antiferromagnet at half filling $n=1$. There are no
phase separation windows in the $J/t \rightarrow 0$ 
limit and all transitions are second order.

For $J/t \gtrsim 1$ the {\it sequence} of magnetic phases,
with increasing filling, remains the same as at weak 
coupling but the window of
spiral states shrink yielding to the FM state at low density
and a window of phase separation near $n=1$. For $J/t \gg 1$
(not shown in the figure) the only surviving states are
the ferromagnet and the $n=1$ Neel state, separated by a
phase separation window. The system heads towards the
`double exchange' limit.

{\bf (ii)}~{\it Weak attraction $(U \sim t)$:}
On a finite lattice the finite size gap $\sim t/L^2$ (in 2D) 
makes it difficult to stabilize a superconducting state below
a $L$ dependent scale.  Since we are using a real space
framework, to connect up with finite $T$ Monte Carlo
calculations later, we have only limited data at $U < 2t$.
Fig.3(b) shows results at $U=2t$ as typical of
`weak coupling'.

At $U=2t$ and $J=0$ we have the usual ${\bf k},\uparrow,~-{\bf k}, 
\downarrow$
pairing. 
At finite $J$ one would (a)~expect the magnetic order to be modified
since the effective spin-spin interaction is now in a finite
$\Delta_0$ background, and (b)~the superconductivity to be weakened
since the pairing is no longer between ${\bf k},\uparrow,~-{\bf k}, 
\downarrow$ but the states 
${\bf k},{\uparrow}$ and $ - {\bf k} + {\bf Q},{\downarrow}$, 
where ${\bf Q}$ is the magnetic ordering vector. 

The first effect is weak since the maximum $\Delta_0 \sim 0.4t$,
opening  only a modest gap in the density of states with 
limited impact on the spin-spin interaction. So the magnetic
character {\it within} the superconducting window, Fig.3(b),
is very similar to the $U=0$ case. The $\Delta_0$ however
falls with increasing $J$, surviving to a scale $J_c(n)$ 
shown in the panel. The maximum of  $J_c$ occurs at $n \sim 0.8$
and the value at $n=1$ is lower than that. 
In the regime $\Delta_0=0$ the magnetic phases are of course as
in panel (a).

{\bf (iii)}~{\it Intermediate attraction  $(U \gg t)$:}
Panel (c) shows data at $U=4t$ and the $\Delta_0 $ at 
$n \sim 0.8$ is now $1.4t$, much larger than at $U=2t$.
As a result, the electronic density of states is
modified with respect to its band character over a wide 
energy window. 

% ----------------------------------------------------
\begin{figure*}[t]
 \centerline{
\includegraphics[height=5.3cm,width=5.5cm,angle=0]{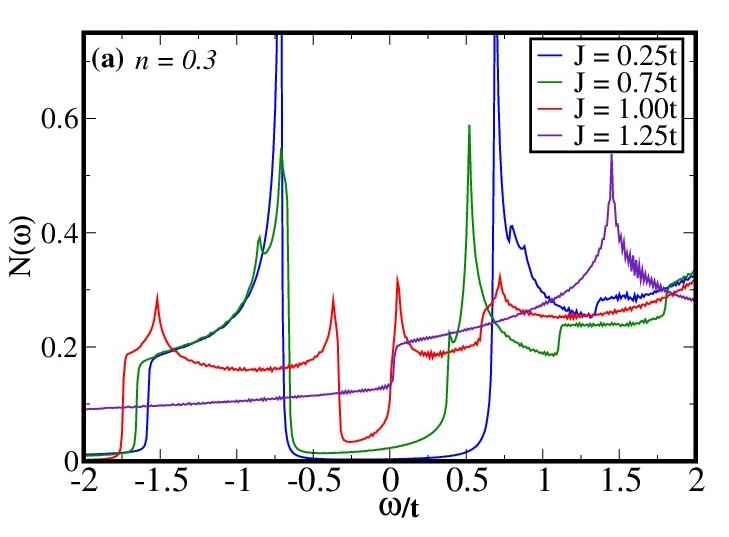}
\includegraphics[height=5.3cm,width=5.5cm,angle=0]{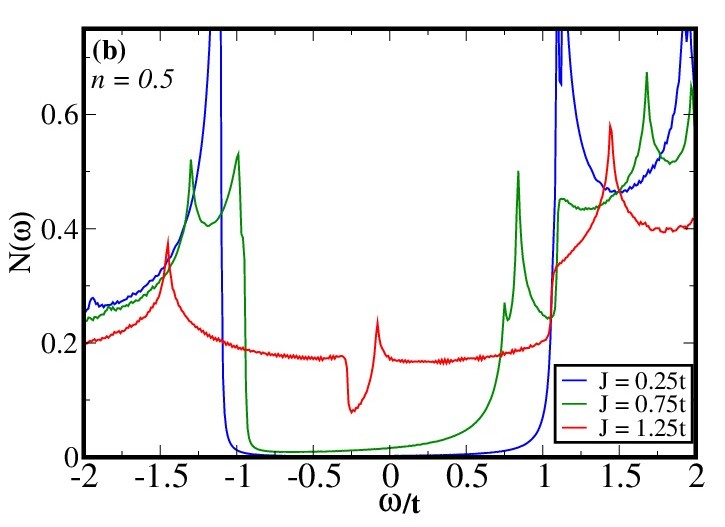}
\includegraphics[height=5.3cm,width=5.5cm,angle=0]{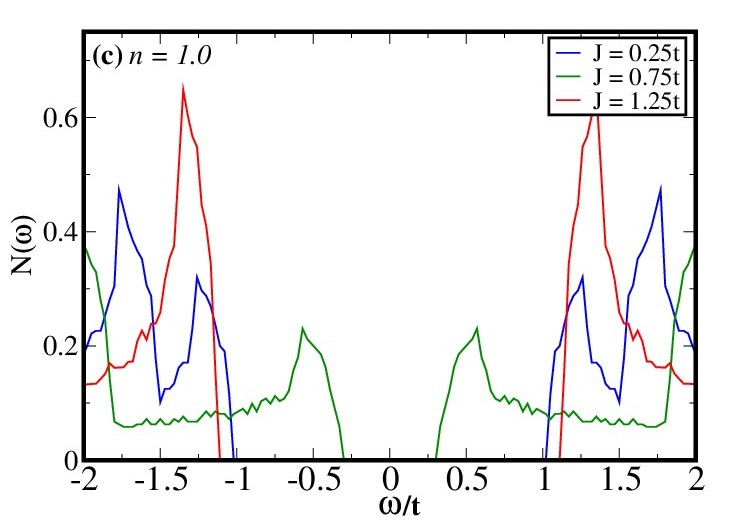}
}
\caption{Color online:
Electronic density of states
at different filling and magnetic coupling at $U=4t$,
on magnetic-superconducting
backgrounds obtained through the variational scheme.
For $n=0.3$ (panel (a)) and $n=0.5$ (panel (b))
the DOS shows transition from a
gapped to a gapless superconducting
state at some  coupling $J_g(n)$.
At $n=1$ the system remains gapped throughout,
however, there is a nonmonotonicity in the behavior
of the gap as one transits from the magnetic
superconductor to the
magnetic insulator at a critical value $J_{c} \sim 0.75t$.
}
\end{figure*}
% ----------------------------------------------------

The changed density of states changes
the spin-spin coupling  and the magnetic phases
show clear differences with respect to the small $U$
cases. These include changes in the magnetic phase
boundaries within the SC phase and the emergence of
a window of Neel order with ${\bf Q} = (\pi, \pi)$,  
close to $n=1$. 

Superconducting order survives 
over a wider range of magnetic coupling with the maximum
$J_c$ being $\sim 1.5t$, occurring at $n \sim 0.6$. Beyond
$n \sim 0.6$ there is a quick drop in $J_c$
as a phase separation window intervenes. The $J_c$ 
at $n=1$ is $\sim 0.75t$, well below the maximum 
at $n \sim 0.6$.

\subsection{Quasiparticle properties}
The magnetic superconducting state involves a suppression
of $\Delta_0$ as $J$ increases. Had the
pairing been between the usual $\vert {\bf k} \uparrow \rangle$
and $\vert - {\bf k} \downarrow \rangle$ states it would
have led to a suppressed BCS gap with the overall character
of the density of states (DOS) remaining unchanged. However,
the pairing now takes place in a magnetic background, where the
Bloch states are superposition of $\vert {\bf k} \uparrow \rangle$
and $\vert {\bf k} + {\bf Q}  \downarrow \rangle$. The {\it combination
} of pairing
and magnetic interaction now connect a larger set of states.
For example $\vert {\bf k} \uparrow \rangle$
connects to $\vert {\bf k} + {\bf Q}  \downarrow \rangle$,
$\vert - {\bf k} - {\bf Q}  \uparrow \rangle$, and 
$\vert - {\bf k} \downarrow \rangle$.
The eigenspectrum that emerges need no longer look like
the `BCS' result. In the section below we describe
the features that we observe and in the section
after we try to analyze these features in terms of the
approximate Green's function theory.

%-----------------------------------------------------
\begin{figure}[b]
\centerline{
\includegraphics[height=5cm,width=4.5cm,angle=0]{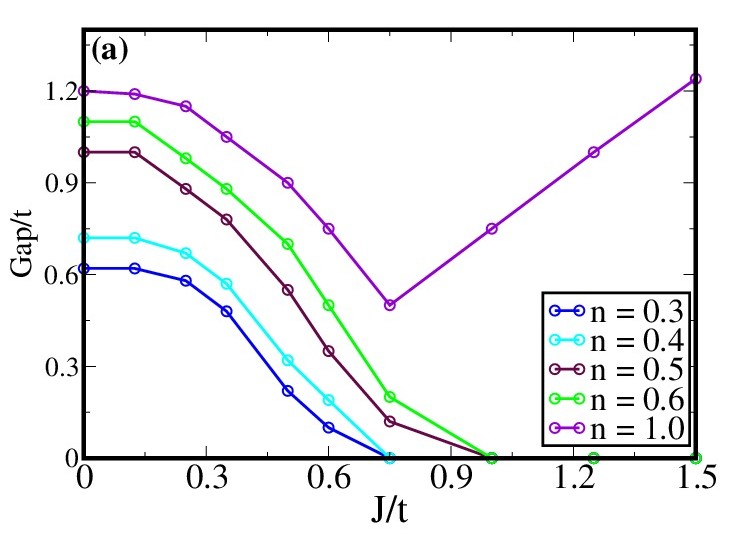}
\includegraphics[height=5cm,width=4.5cm,angle=0]{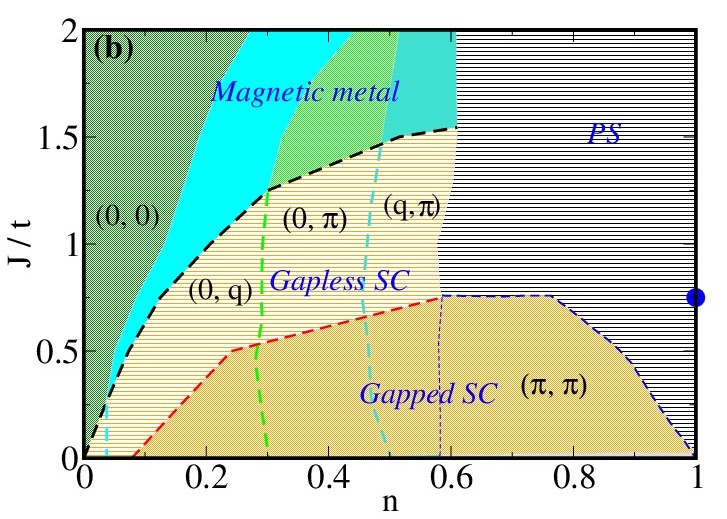}
}
\caption{Color online:
(a)~Gap in the DOS plotted as a function of magnetic
coupling for different fillings. At $n=1$ for $J \le 0.9t$ the
 superconducting gap gets progressively suppressed with
 $J$. Beyond $J \sim 0.9t$ the gap is the antiferromagnetic
 gap which increases with $J$. At $n \ne 1$, the gap reduces
 monotonically with $J$, in agreement with $\Delta_{0}$
(see Fig.2a).  (b)~$n-J$ phase diagram at $U=4t$ showing
 the gapped and gapless superconducting phases.
}
\end{figure}
%-------------------------------------------------------

\subsubsection{Density of states}

Fig.4 shows the electronic DOS computed on backgrounds obtained through
the Green's function calculation.  The three panels comprise of DOS 
pertaining to three density regimes and varying $J$.  The attractive 
interaction is $U=4t$ in all cases.

%---------------------------------------------------------------
\begin{figure*}[t]
\includegraphics[height=12cm,width=12cm,angle=0]{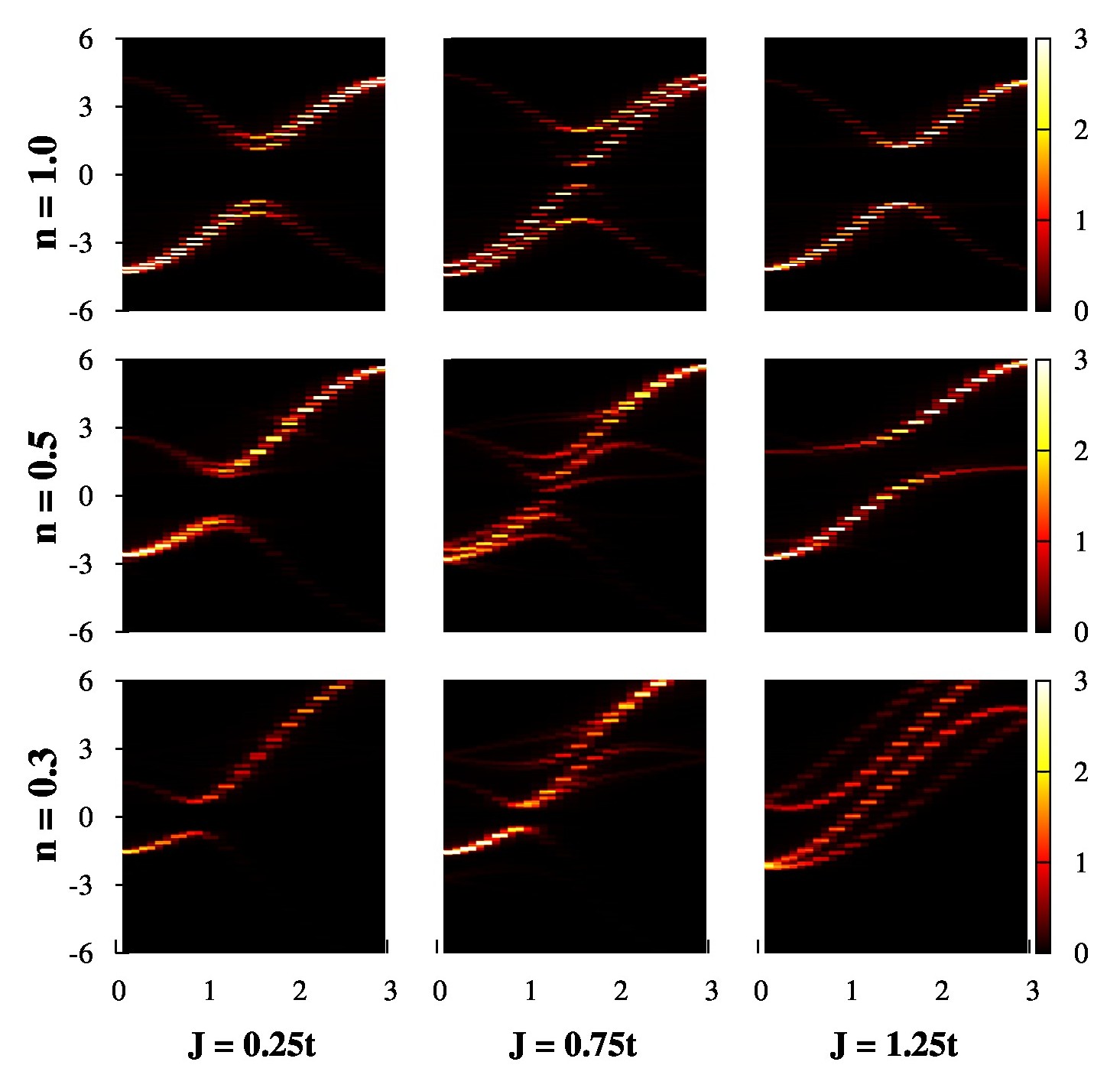}
\caption{Color online: The spin summed electron spectral function,
$A({\bf k},\omega)$ for ${\bf k}$ varying from $(0,0) $ to $(\pi, \pi)$
at different combinations of $n$ and $J$ and $U=4t$.  At $n=1$ (top
row) the gap near $(\pi/2,\pi/2)$ reduces from $J=0.25t$ to $J=0.75t$
but increases again at larger $J$. There are also multiple bands visible
at $J=0.25t,~0.75t$.  At $n = 0.5$ and $n = 0.3$ the low $J$ result is
almost BCS like, with only two bands visible, while the $J=0.75t$ case
shows a large number of bands, with one crossing $\omega=0$. At larger
$J$, as $\Delta_0$ becomes very small, the bandstructure simplifies again
and is mostly described by the `magnetic metal' limit.  The results are
shown for $36 \times 36$ lattice.}
\end{figure*}
%---------------------------------------------------------------

{Fig.4(a) shows the situation at filling  $n = 0.3$.
The spectrum remains gapped at weak $J = 0.25t$ (modulo a
`tail' due to the lorentzian broadening) and has the usual
gap edge singularities  akin to the $J=0$ case.
At $J=0.75t$, however, there is finite DOS at $\omega=0$
and the remnant of the  `gap edges' have moved inward.
The inward movement of the edges can be attributed to the
reduced $\Delta_0$ as $J$ increases but the low energy
DOS involves a new band.
$J=t$ shows even larger DOS at $\omega=0$ and makes
visible new van Hove singularities.
The understanding of these features come from an
analysis of the dispersion using the momentum
resolved spectral functions. We take that up in the
next section and just highlight the features in the
changing DOS here.

%---------------------------------------------------------------
\begin{figure}[t]
\includegraphics[height=6cm,width=8.8cm,angle=0]{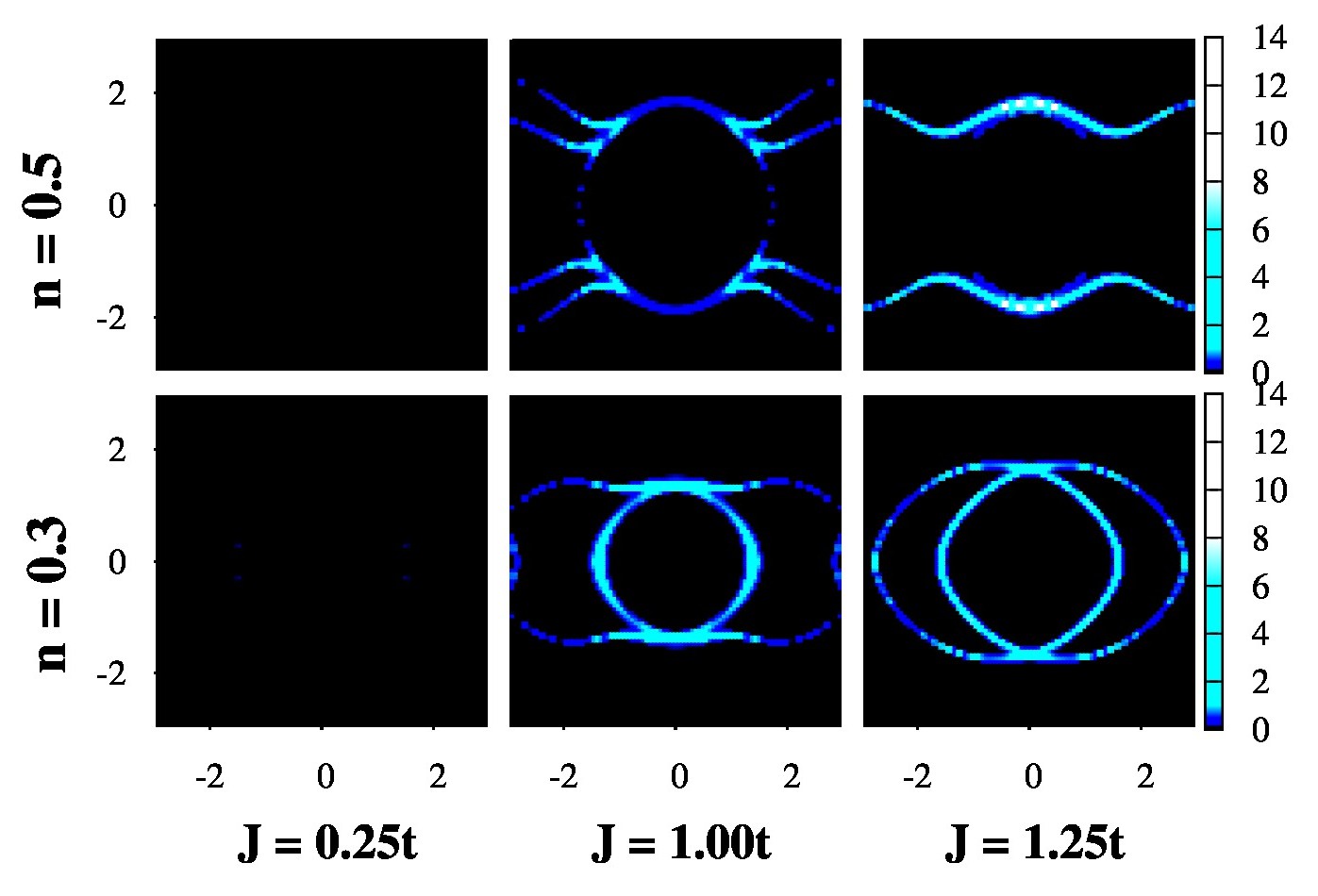}
\caption{Color online: Low energy spectral weight at the Fermi
level for different $n-J$ cross sections. The parameters are the same
as in Fig. 6. An weak $J/t$ give rise to a gapped state
and consequently there is no low energy weight near the Fermi
level. Increase in $J/t$ leads to pile up of spectral weight
near the Fermi level whose symmetry is dictated by the
underlying magnetic wave vector ${\bf Q}$. The distribution
of the spectral weight near the Fermi level is anisotropic,
indicative of a nodal Fermi surface.
}
\end{figure}
%---------------------------------------------------------------

At $n=0.5$ the observations are qualitatively similar
to the $n=0.3$ case, with finite DOS at $\omega=0$
being visible at the two upper values of $J$. The overall
`gap structure' within which the low energy features are
seen is wider at $n=0.5$ due to the larger $\Delta_0$.

{The behavior at $n=1$, Fig.4(c), is distinctly different.
The presence of satellite peaks within the BCS like gap is 
significant in this case. The spectrum is gapped at all magnetic 
coupling but the gap shows nonmonotonic behavior. Initially
increase in 
magnetic coupling pushes the satellite peaks  
to low energy  narrowing the gap. However the pairing amplitude itself
vanishes at a critical $J \sim 0.75t$, beyond which the system changes to
a magnetic insulator - with the gap now being proportional to and 
sustained by $J$.
}

\subsubsection{Gapped and gapless regimes}

Fig.5(a) shows the $J$ dependence of the gap at 
different filling. At weak magnetic coupling 
the superconducting gap follows the behavior of the pairing 
field amplitude and undergoes suppression with increasing 
$J$. At half filling, till a coupling of $J \sim 0.9t$
the behavior of the gap is the same as that of its low filling
counterpart. For  $J \gtrsim 0.9t$ the gap increases linearly with 
$J$. The gap in this regime arises from antiferromagnetic 
($\pi, \pi$) order.
For $n \neq 1$ the gap vanishes at a scale we call $J_g(n)$.

Fig.5(b) shows the $n-J$ phase diagram at $U=4t$, now with 
the superconducting phase demarcated into gapped 
and gapless regimes. The gapped regime is characterized 
by the presence of large $\Delta_0$ while the
gapless window has relatively smaller $\Delta_0$. That by 
itself does not explain why the qualitative character of the
DOS changes, so we examine the electron dispersion in the
magnetic superconductor to explore this issue.

\subsubsection{Electron dispersion}

Fig.6 shows the momentum 
resolved spectral function $A({\bf k}, \omega)
= \sum_{\sigma} A_{\sigma}({\bf k}, \omega)$ for three
different
$n-J$ combinations. The momentum scan is along the
diagonal of the Brillouin zone, ${\bf k} =
(0,0) \rightarrow (\pi,\pi)$.
Since the spectra are computed on an
ordered state there is no broadening of the lines and
we essentially map out the multi-branch dispersion in
the magnetic-superconducting state.

We begin with $n=1$, top row. At weak magnetic coupling,
$J = 0.25t$, the behavior is BCS like with the 
characteristic {\it back bending} feature in the dispersion curves. 
The effective gap is slightly reduced compared 
to its BCS value, and there is a small branching visible
for ${\bf k} \sim (\pi/2,\pi/2)$.
At $J=0.75t$ the branching feature is far more prominent
and the separation between the inner branches, that sets
the gap, is much smaller than at $J=0.25t$. 
${\bf k}$ regions associated with 
${\partial E_{\alpha}({\bf k})}/{\partial {\bf k}}=0$,
where $E_{\alpha}({\bf k})$ are the dispersion,
lead to the van Hove singularities
observed in Fig.4(c).

At $n = 0.5$, middle row, weak $J$ essentially reproduces
the BCS result, with a smaller gap than $n=1$ due to
the smaller $\Delta_0$ - occurring at a lower ${\bf k}$
due to the lower filling.
At $J = 0.75t$ a very complex picture emerges, with in
principle all the 8 bands that arise from BdG being 
visible (although a six band, Green's function based,
approach captures the essential features). Along the
$(0,0) \rightarrow (\pi,\pi)$ scan one of the bands
seems to cross $\omega=0$. The multiple and
prominent $E_{\alpha}({\bf k})$ generate the van Hove
singularity structure seen in Fig.4(b). At $J=1.25t$
the $\Delta_0$ is very small and the features
are similar to that of a magnetic metal.

At $n = 0.3$ the qualitative features are similar to
$n=0.5$ although the multiple bands are not all visible
for the color scheme that we have used.
The superconducting state survives to $J_c \sim t$ and the
$J =1.25t$ result is for a magnetic metal.

While it is difficult to extract useful analytic expressions
for the three branches of the dispersion from each
$G_{\sigma\sigma}({\bf k}, \omega)$, explicit
functional forms can be obtained in the gapless phase
when $\Delta_0 \lesssim J$. We provide these results in
the Appendix, and have cross checked them with respect
to the numerical results.

\subsubsection{Low energy weight distribution}

In connection to the spectral features discussed above in Fig.7
we show the distribution of low energy spectral weight 
across the Brillouin zone at low and intermediate filling (at
$n=1$ the spectrum is always gapped). At weak 
magnetic coupling the spectrum is gapped out and thus there is 
no low energy weight. 

We computed the ${\bf k}$ dependent spectral weight at
$\omega =0$,
summed over spin channels, $A({\bf k}, 0) = \sum_{\sigma}
A_{\sigma\sigma}({\bf k}, \omega)$, where: 
\begin{eqnarray}
A_{\uparrow \uparrow}({\bf k}, 0)
& =& -(1/\pi) Im~{1 \over {i \eta - (\epsilon({\bf k}) -\mu) 
- \Sigma_{\uparrow \uparrow}({\bf k}, i \eta)}} 
\vert_{\eta \rightarrow 0} \cr
~~&&\cr
~~&&\cr
\Sigma_{\uparrow \uparrow}({\bf k}, i\eta) & = &
{\Delta_0^2 \over {i \eta  + (\epsilon({\bf k}) -\mu)}}
+ {J^2 \over {i \eta - (\epsilon({\bf k} + {\bf Q}) -\mu)}}
\nonumber
\end{eqnarray}
%
% extract an analytic form for the low E wt..
%
etc. The results in Fig.7 highlight the rather strange
looking `Fermi surface' that emerge. The low $J$ panels
show no spectral weight since the system is gapped. 
$J=t$ shows non trivial Fermi surfaces in the superconductor,
dictated by the magnetic wavevector, while
$J=1.25t$ is superconducting for $n=0.5$ and 
a magnetic metal for $n=0.3$.

\section{Discussion}

This section covers some issues of method, related to the approximations
that we have made in handling the model in Eqn.1, and the phase diagram,
in terms of the magnetic coupling and attractive interaction. We
comment on what it suggests for spectral features in the borocarbides.

\subsection{Computational issues}

\subsubsection{The `classical' approximations}

The model in Eqn.1 involves an attractive electron-electron
interaction $U$ and the coupling $J$ between the electron
spin and a local moment of spin $S$. This describes
interactions between quantum degrees of freedom, and, beyond
weak coupling, is very non trivial. 
The treatment of the Hubbard interaction in terms of a
classical pairing field, and of the spin $S$ as
classical, makes the model tractable by reducing it to
a variational problem determining a static $\{\Delta_i, {\bf S}_i\}$
background that minimizes the electron energy.

% -------------------------------------------------------------------
\begin{figure}[b]
\includegraphics[height=8.0cm,width=9.0cm,angle=0]{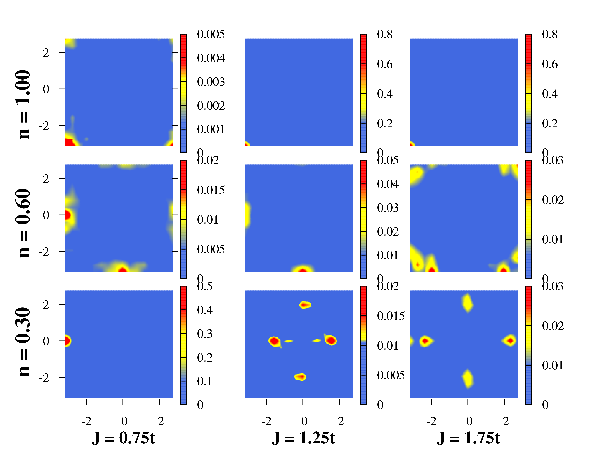}
\caption{Color online:
Magnetic structure factor at $T \sim 0$ for different filling and 
magnetic interaction $J$.  At half filling ($n = 1$) the MC always 
leads to a ${\bf Q} = (\pi,\pi)$, Neel, state, as in the VC.  At the 
intermediate filling of $n = 0.6$ a $(0, \pi)$ and a $(q, \pi)$ state 
is realized for the particular choice of the magnetic coupling, in 
agreement with the VC results. At low filling of $n = 0.3$ and 
intermediate and strong magnetic coupling the state as obtained 
through MC slightly deviates from that obtained through the VC, with
the $(0, q)$ being now replaced by $(0, \pi)$, the neighboring
phase in the VC phase diagram.
}
\end{figure}
% -------------------------------------------------------------------

The mean field approximation for $U$ makes qualitative
sense as long as $\Delta_0 \neq 0$.
The presence of superconducting 
order at $J=0$ is well known, the persistence of order at small
$J$ has also been established via numerically exact methods. This
suggests that the mean field treatment of $U$  is a valid
first approximation. 
Quantum fluctuations of the pairing field would be important
near $J_c$ in the large $U$ problem, where the mean field
amplitude vanishes, but correlation effects would be
significant. We have not focused on that regime here.

The treatment of the local moment as
`classical' is valid when $2S \gg 1$. For the borocarbides
$4f$ shells for the magnetic superconductors involve $2S \sim 3-5$
and the classical treatment again ought to be reasonable.
There are, however, low moment, and non magnetic, superconductors
involving Tm and Lu which {\it cannot} be captured well 
within our scheme. 

\subsubsection{Single -vs-  multichannel decomposition of interaction}

We have considered the effect of $U$ only in the pairing 
channel,
and the magnetic response arises from the ${\bf S}_i$.
As a first approximation this is justified because the 
pairing and magnetic effects arise from different couplings in
our model (the $U$ is not primarily responsible for the
magnetic order). However, there would be a renormalisation
in the magnetic sector arising from the $U$, if we were to
consider an additional magnetic decoupling of the Hubbard
term. We discuss this below.

% -------------------------------------------------------------------
\begin{figure}[b]
 \centerline{
 \includegraphics[height=5cm,width=4.5cm,angle=0]{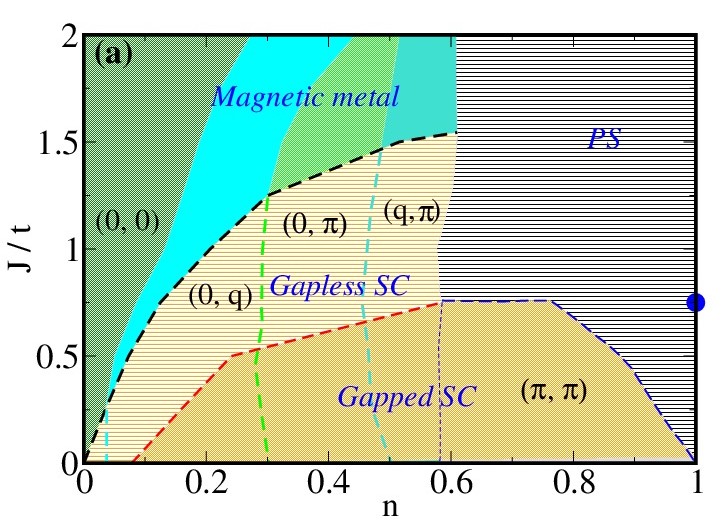}
 \includegraphics[height=5cm,width=4.5cm,angle=0]{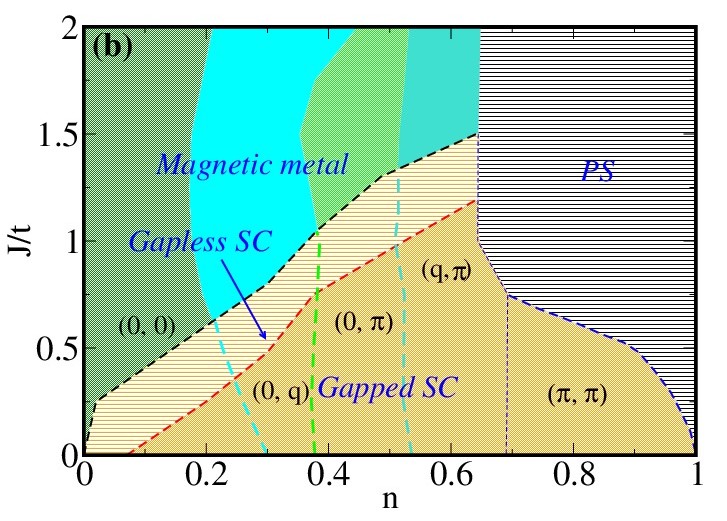}
 }
 \caption{Color online:
The ground state $n-J$ phase diagram as obtained
 through MC (right) in comparison to the one obtained through
 the variational calculation (left) at $U = 4t$. Notice that
 the gapless regime shrinks in the MC phase diagram as compared
 to the one obtained through VC. The emergence of the Neel,
 $(\pi, \pi)$ antiferromagnetic window near $n = 1$ is verified
 through the MC as well.}
\end{figure}
% -------------------------------------------------------------------

Decomposing $U$ in both the magnetic and pairing channels lead to the 
effective Hamiltonian,
\begin{eqnarray}
H & = & H_{kin} + H_{pair} 
- \sum_i \{(JS_i^+ - h_i^+) \sigma_i^- + h.c \} \cr
&& ~~~~~~~~~~~~~~~~~~~~~~~~+ \vert U \vert
\sum_i \{ \vert \Delta_i \vert^2 +  
\langle \sigma_i^{+} \rangle \langle \sigma_i^{-} \rangle
\} 
\nonumber
\end{eqnarray}
where, $h_{i}^{+} = U\langle \sigma_{i}^{+}\rangle$ and 
$\sigma_{i}^{+} = c_{i\uparrow}^{\dagger}c_{i\downarrow}$, 
{\it  etc}.
For $h_i^+ = U \langle \sigma_i^+ \rangle $ to
be nonzero 
does not require symmetry breaking driven by $U$. There is
a `source term',  
{\it since $J S_i^-$ already forces $\langle \sigma_i^{+} \rangle
\neq 0$}. So, the leading effect of the magnetic decoupling 
can be estimated simply by calculating
$  U \langle \sigma_i^+ \rangle_0$, where the 
subscript zero refers to the model with only pairing 
decomposition. 

We have checked that the `original' exchange field $J S_i^+$
and the renormalised field 
$JS_i^+ - U \langle \sigma_i^+ \rangle_0 $ have the same
spatial character, so the leading effect of the magnetic
channel can be included via a renormalisation $J 
\rightarrow J_{eff}$. 
The effective exchange field is {\it smaller} than
the bare field by $15-20\%$, which we think arises
due to the 
diamagnetic tendency of the attractive $U$ term. The weaker
$J_{eff}$ will expand
the domain of superconducting order marginally without 
affecting any qualitative conclusion.

\subsubsection{Comparison with unrestricted minimization}

Fig.8 shows the magnetic structure factor computed at
different filling for three different regimes of the
magnetic interaction. In the intermediate and strong
coupling regimes, cooling down the system from
an uncorrelated high temperature state reproduces the
magnetic order as has been obtained through the
variational calculations. In the weak interaction
regime however, the system fails to attain the
global minimum in the energy landscape within
the limited annealing time and finite system size. The
configuration thus obtained through the Monte
Carlo is often energetically unfavorable compared
to the one obtained variationally.
Nevertheless, over a wide parameter space the
variational ground state is well reproduced on cooling
down from a high temperature state.

The resulting ground state phase diagram
is shown in Fig.9, in comparison to the one obtained
through the variational scheme. The ground state as
obtained through the Monte Carlo certainly agrees
qualitatively with all features of the
variational result, and also confirms that the
`homogeneous' $\Delta_i$ assumption for the
ground state is not unreasonable.

\subsubsection{Coexistence of modulated pairing order 
with ferromagnetism}

Our variational calculation suggests that a homogeneous
superconducting state cannot coexist with a large ferromagnetic
internal field $JS$. However, it is known 
\cite{fflo_assort, mpk_bp} that homogeneous superconducting
order {\it can exist} in the presence of a weak external 
magnetic field, beyond which there is a narrow regime of
modulated Fulde-Ferrell-Larkin-Ovchinnikov (FFLO)
order, before pairing is lost.
This effect does exist in
our phase diagram as well, but over a very narrow window so
it has not been given prominence in Fig.3. We comment on
this below.

For a superconductor in an applied
field $h$, the FFLO state exists over a window $h_1(n)$ 
to $h_2(n)$ \cite{fflo_assort, mpk_bp}.  Below $h_1$ the system remains 
a homogeneous
superconductor, with zero spin polarization.
This is traditionally called the `unpolarised superfluid'
(USF) state.  Above $h_2$ the system is a magnetized
normal Fermi liquid. The equivalent in our model are two
magnetic couplings $J_1(n)$ and $J_2(n)$. Knowing $h_{1}(n)$ 
and $h_{2}(n)$ one can just superpose these on the ferromagnetic 
window of the $n-J$ phase diagram to locate the USF and
FFLO regimes. Fig.3 shows these tiny windows, virtually 
invisible at $U=2t$.
The reason the window is so small is due to the 
tiny density window over which ferromagnetism shows up at
small $J$, and the {\it small $J_1$ and $J_2$ scales
in the small $n$ window}. $J_1$ and $J_2$ are related to
the pairing gap in the spectrum, and this vanishes as
$n \rightarrow 0$.

In summary, a local moment polarized homogeneous 
superconductor, and a pair modulated ferromagnetic
state, can exist in our model, but over a tiny density
and $J$ window.

\subsubsection{Size limitations}

The variational calculation, when cast in momentum space,
does not have significant size limitations, except in the
number of ${\bf q}$ values over which the energy has
to be minimized. 

A more serious size limitation arises when 
Monte Carlo based simulated 
annealing is used for `unrestricted' minimization, and 
for accessing finite temperature properties.
This requires iterative diagonalization of a 
$4N \times 4N$ matrix (where $N = L^2$) and even when a
cluster algorithm is used for the MC updates only sizes
upto $24 \times 24$ can be accessed within reasonable
time. We have checked that thermodynamic properties can
be accessed down to $U=2t$ reliably on these sizes, but
the subtle spectral features that one observes in the
large size ground state calculations cannot be 
resolved well on these sizes. We also cannot go
down to $U \sim t$, which we believe is appropriate
for quantitative description of the borocarbides.

\subsubsection{Benchmarking the Green's function results}

The BdG problem generates 8 bands for a given ${\bf k}$
since each $\vert {\bf k} \sigma \rangle $ connects to
three other states via pairing and magnetic scattering,
and the results for $\sigma = \uparrow$ and $\sigma = \downarrow$ 
are now non degenerate.  Some of the residues associated with
these bands can, however, be quite small and hard to
identify. The Green's function approach on the other
hand truncates the scattering processes to
$J^2$ and $\Delta_0^2$, dropping $J^2 \Delta_0^2$, and
the resulting Green's function has three poles for
each ${\bf k} \sigma$. 
The results are obviously exact at $J=0$ or $\Delta_0 =0$,
but, as the results in Fig.10 show, they are surprisingly
accurate over a large $\Delta_0-J-n$ parameter range.

The results however are {\it not} accurate for the
magnetic superconductor at $n=1$ where
an unusual DOS emerges (see Fig.4(c)) and also
at low $J$ at other densities where a spurious
low energy band with a small residue, $\propto J^2$,
emerges. Away from these parameters the Green's function
approach provides a useful tool for understanding
the complex band structure.

% -------------------------------------------------------------------
\begin{figure}[b]
 \centerline{
 \includegraphics[height=11.5cm,width=8.5cm,angle=0]{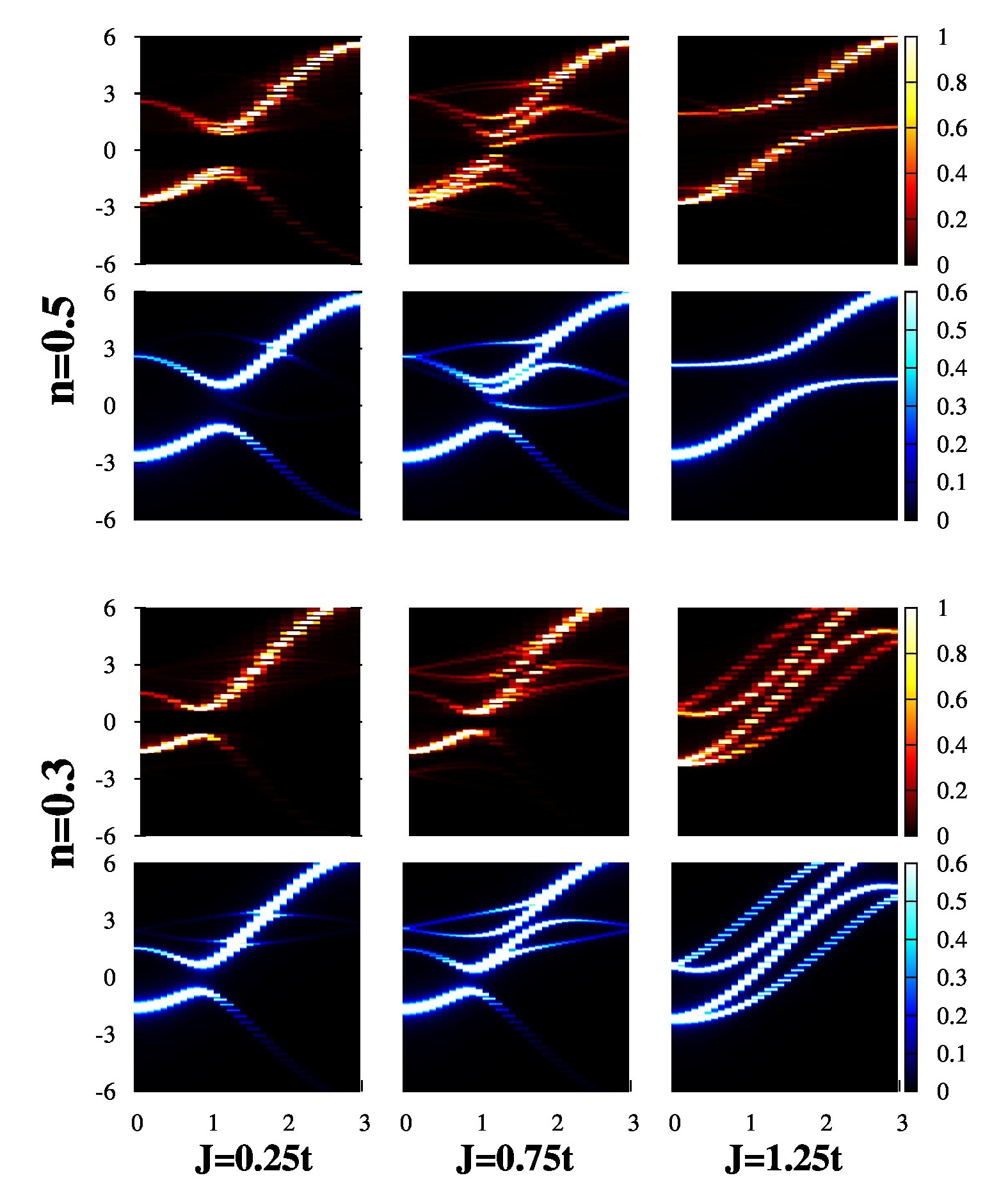}
 }
\caption{Color online:
Comparison of the total spectral function $A({\bf k}, \omega)$
obtained from the BdG diagonalization and
the Green's function method, for parameters
mentioned in the figure. The top row in each set corresponds to the
BdG results while the bottom row shows the Green's function
result. The agreement is reasonable for all the parameters shown
here.
The results here are at $U=4t$ and for momenta discretised on a
$36 \times 36$ lattice.}
\end{figure}
% -------------------------------------------------------------------
% -------------------------------------------------------------------
\begin{figure}[t]
\centerline{
\includegraphics[height=4cm,width=4.3cm,angle=0]{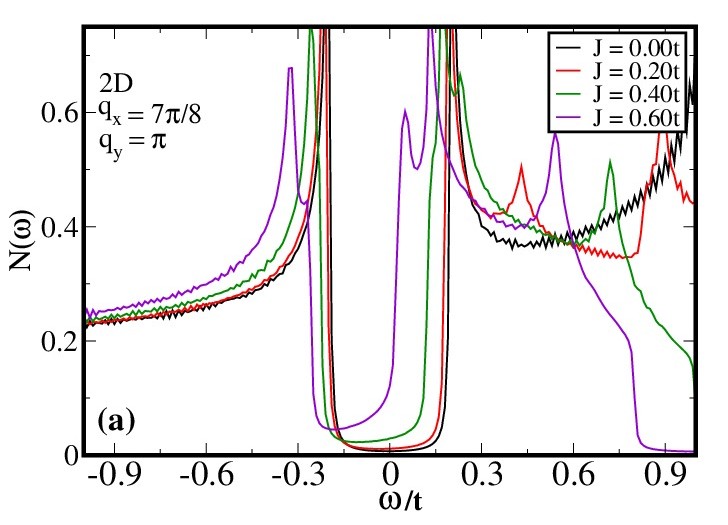}
\includegraphics[height=4cm,width=4.3cm,angle=0]{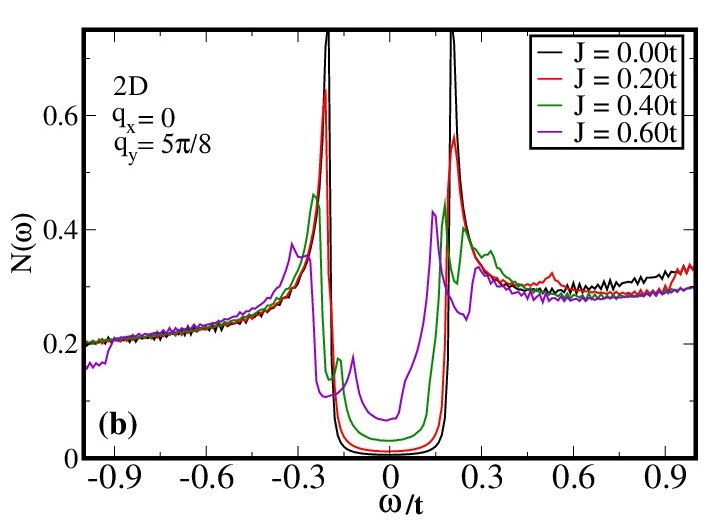}
}
\centerline{
\includegraphics[height=4cm,width=4.3cm,angle=0]{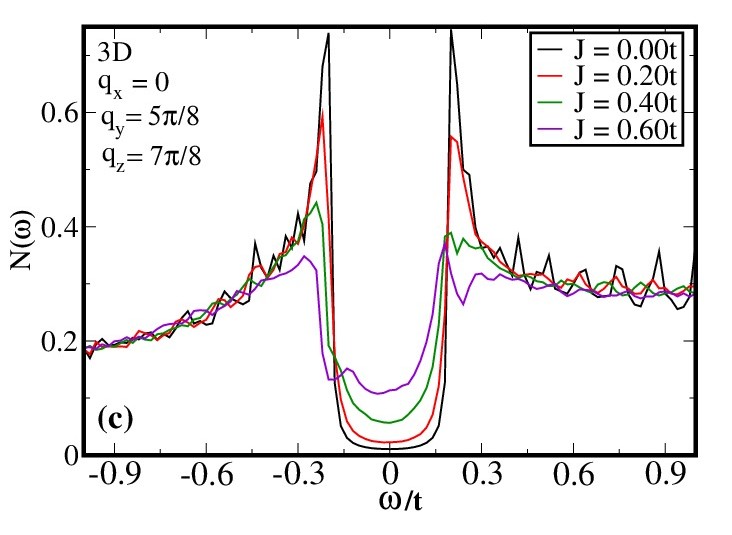}
\includegraphics[height=4cm,width=4.3cm,angle=0]{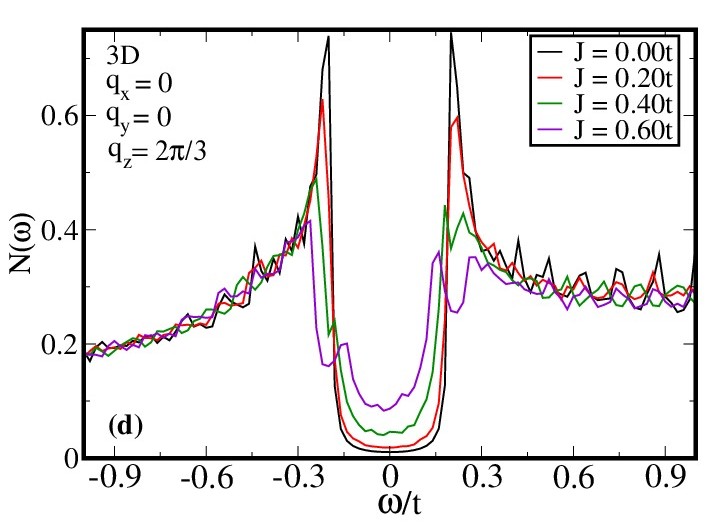}
}
\caption{Color online:
Top row: Electronic density of states for the
square lattice on two different magnetic backgrounds and
varying $J$, at $n = 0.5$.
Bottom row:
Same as above for a three dimensional (cubic) lattice, for the
choice of the magnetic wave vectors shown in the individual panels.
The DOS are computed through the Green's function method 
for a pairing field amplitude of $\Delta_{0} = 0.2t$, plausible
for a weak interaction regime. In both 2D and 3D the
system undergoes transition from a gapped to a gapless state
with increasing magnetic coupling.}
\end{figure}
%---------------------------------------------------------------------

\subsubsection{Extension to low $U/t$}

Since the thermal physics cannot be worked out on lattices
beyond a certain size (say with $N_{max} \sim 30 \times 30$)
we have restricted our study mainly to $U \gtrsim 2t$. However
it is worth exploring if a gapless superconducting phase 
can  arise at much lower $U$, and therefore much smaller
$\Delta_0$, than we have studied till now. This will be
relevant for real materials which are mainly in the 
weak coupling, $U \lesssim t$, limit.
In Fig.11(a) and 11(b) we show the DOS  calculated through 
the Green's function method for typical spiral magnetic 
backgrounds, with  ${\bf Q}$ marked in the Fig,
and pairing field amplitude set to
$\Delta_{0} = 0.2t$. We study both the two dimensional and
three dimensional case (which is experimentally more
relevant) and find that all cases show a gapped to
gapless transition with increasing $J$ on a scale
$J_g \sim \Delta_0$. 

This little demonstration is just meant to emphasize
that the occurrence of a gapless phase at finite $\Delta_0$
is not an artifact of large $U$ or two dimensionality
and can well occur in weak coupling 3D superconductors
as well.

\subsection{Relating to experiments}

\subsubsection{$U-J$ phase diagram}

The results at $U = 4t$ are part of a larger $U-J-\mu$ 
phase diagram.
In real solids the attractive interaction would be
typically much smaller that $4t$ (and in
possible cold atomic systems they could be larger).
Keeping this in mind we attempted to map out the $U-J$
phase diagram at a few densities.
Fig.12, top row,
 shows our results at $n \sim 0.5$ and $n \sim 0.3$.

We find the following:
(i)~At $n \sim 0.5$ over the range of
$U$ the system exhibits G-type antiferromagnetic order ($\pi, \pi$)
or ($q, \pi$) order depending upon $J$.
The superconducting phase makes a gapped to gapless transition
at a $J$ that increases with $U$.
(ii)~At $n = 0.3$ the magnetic state can be
($\pi, \pi$),
($q, \pi$) or ($0, \pi$). The superconducting state is gapped
or gapless depending upon the strength of the magnetic interaction,
with the large $U \gg J$ regime favoring gapped superconductivity.
The $J \gtrsim U$ regime
again gives rise to gapless SC and finally a magnetic metal.
No phase separated regime is realized at low filling.

%---------------------------------------------------------------
\begin{figure}[t]
\centerline{
\includegraphics[height=5.2cm,width=4.4cm,angle=0]{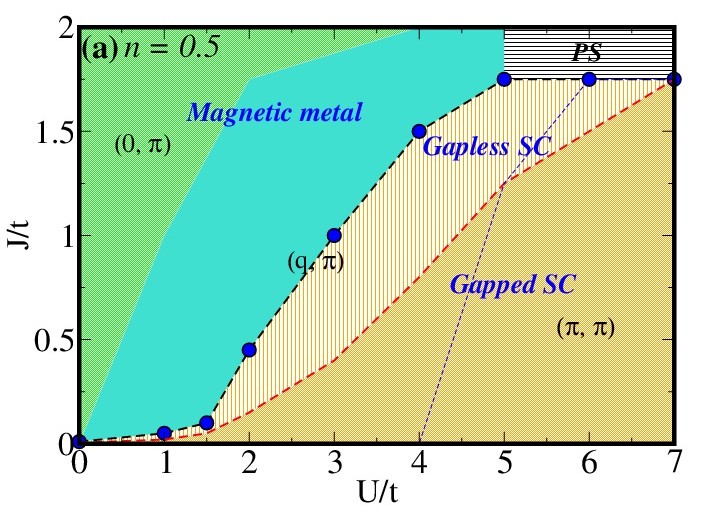}
\includegraphics[height=5.2cm,width=4.4cm,angle=0]{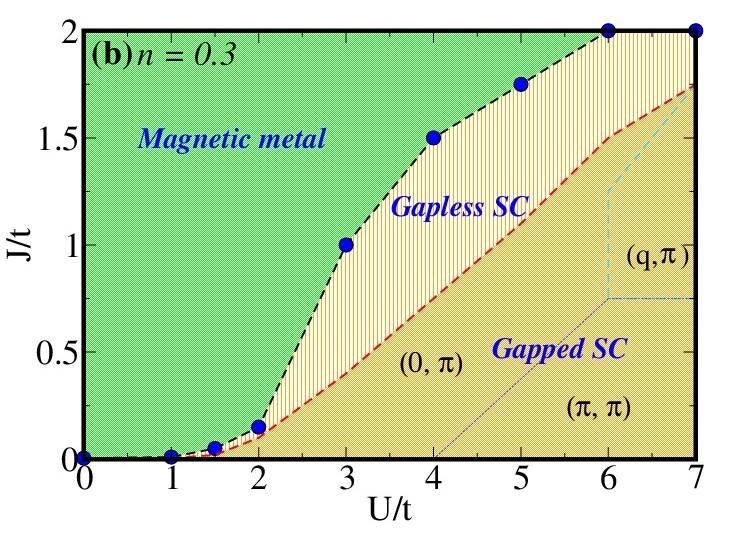}
}
\centerline{
\includegraphics[height=5.2cm,width=4.4cm,angle=0]{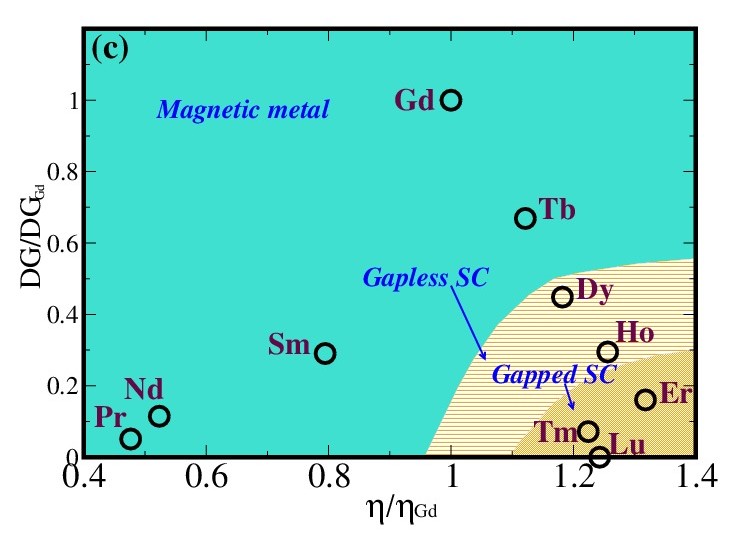}
\includegraphics[height=5.2cm,width=4.4cm,angle=0]{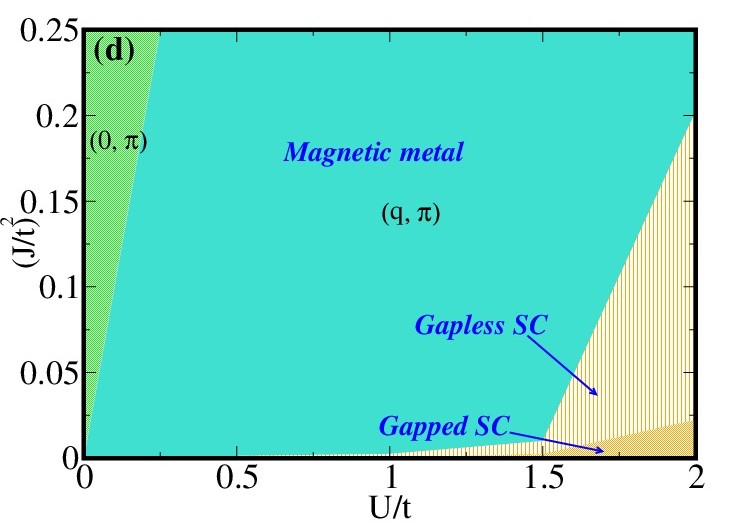}
}
\caption{Color online: 
(a)-(b) Ground state $U-J$ phase diagram at
$n \sim 0.50$ (panel (a)) and  $n \sim 0.3$ (panel (b)).
(c)~Organization of the borocarbide ground state in terms
of the de Gennes (DG) factor, ${S(S+1)}$, denoting the
strength of local electron-spin coupling, and the Hopfield
parameter, $\eta$, indicative of the strength of pairing interaction.
The DG factor is like our $(JS)^2$, while $\eta$ relates to $U$.
}
\end{figure}
% ----------------------------------------------------------------

\subsubsection{The borocarbide phase diagram}

In the borocarbides neutron scattering experiments reveal 
the nature of magnetic order. There is an overall
similarity in the order as one goes down
from GdNi$_{2}$B$_{2}$C, where the DG factor is largest,
to TmNi$_{2}$B$_{2}$C, where the DG factor is smallest,
through  DyNi$_{2}$B$_{2}$C,  HoNi$_{2}$B$_{2}$C, {\it etc}.
All of them seem to have a ${\bf Q} = (0,0,q)$ pattern
of ordering, with $q \sim 0.55$. The order is 
ferromagnetic  in the basal plane with a spiral along
the $c$-axis \cite{lynn1997}.  

Two material parameters are believed to be important in these 
compounds. They are (i)~the  de-Gennes factor (DG) that we have 
already introduced, proportional roughly to our $(JS)^2$, and 
(ii)~the `Hopfield parameter', $\eta$, defined \cite{michor2000}  
as, $\eta = N(E_{F})\langle I_{\alpha}^{2}\rangle$, 
where $\langle I_{\alpha}^{2}\rangle$ is average of 
the electron-phonon matrix element over the atoms 
R, Ni, B, C, and $N(E_F)$ is the DOS
at the Fermi level. $\eta$ relates  roughly
to $U$ in our case.  Both DG and $\eta$ 
have been tabulated for the borocarbides.
% ----------------------------------------------------------------------
\begin{figure}[b]
 \centerline{
 \includegraphics[height=6.0cm,width=6.5cm,angle=0]{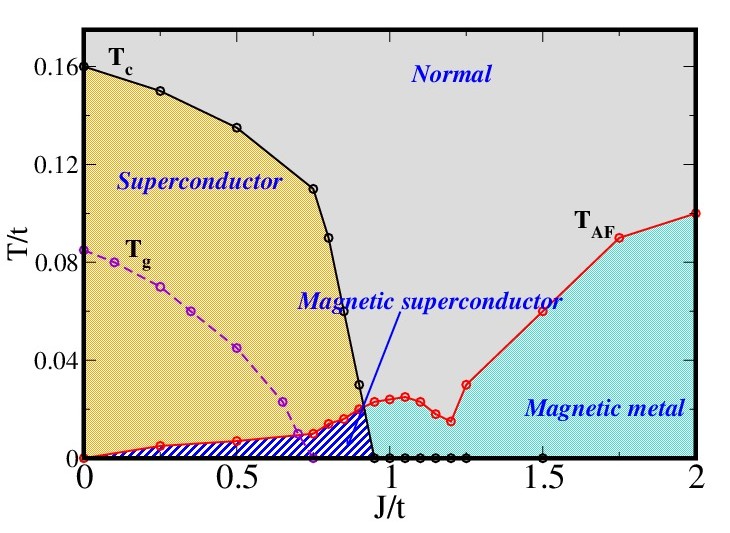}
 }
 \caption{Color online: 
$J-T$ phase diagram at  $n = 0.5$.
 Magnetic coupling strongly suppresses the superconducting $T_{c}$
 and at intermediate coupling there is coexistence of the magnetic
 and superconducting order. In addition to the superconducting
 $T_{c}$ we show a magnetic scale $T_{AF}$ below which the correlation
length grows exponentially (but there is no true long range order
since we are in two dimensions). A third scale,
 $T_{g}$, showing the transition from a gapped 
 to a gapless superconducting phase, emerges.}
\end{figure}
% ----------------------------------------------------------------------
We organized the experimental phase diagram of the borocarbides
in terms of $\eta$ and DG (normalizing by the value for Gd),
Fig.12.(c), and compare it with our variational $U-J$ phase
diagram at a typical density $(n=0.5)$ in Fig.12.(d).
For the real materials the magnetic and superconducting 
boundaries are well established but the possible 
`gapped to gapless' boundary that we show in 12.(c) is our 
conjecture based on 12.(d). 
We believe that 
unless the `multiband' character of the real materials
invalidates the basic picture there must be
an increase in the gap anisotropy (if not a gapless
state) as one moves to increasing DG from
Er $\rightarrow$ Ho $\rightarrow$ Dy, before 
superconductivity is lost in Tb.
There is indeed some evidence for gap anisotropy
and nodal quasiparticles
in the borocarbides, we review that quickly below.

The borocarbides studied involve two `non magnetic' 
compounds, YNi$_{2}$B$_{2}$C and LuNi$_{2}$B$_{2}$C (which
do not have local moments), in addition to those with finite
$4f$ moment and DG factor.
Among both the non magnetic and magnetic 
superconductors one observes an apparent 
direction dependence of the gap 
on the Fermi surface \cite{canfield2003,izawa2002,takagi2004,shin2010, 
baba2008,rybaltchencho1999,yanson2008,schultz2011}. 

In case of YNi$_{2}$B$_{2}$C and
LuNi$_{2}$B$_{2}$C scanning tunneling spectroscopy  (STS)
\cite{canfield2003}, c-axis thermal conductivity \cite{izawa2002}, 
ultrasound attenuation \cite{takagi2004}, {\it  etc.},
suggest that the superconducting 
energy gap  is of the anisotropic s-wave type, 
with point nodes along [100] and [010]. 
The tunneling current in STS  \cite{canfield2003}
as well as the $\sqrt{H}$ dependence of finite field 
heat capacity \cite{izawa2002} suggests the 
presence of low energy quasiparticles.
Angle resolved photoemmission spectroscopy (ARPES) 
on YNi$_{2}$B$_{2}$C suggests \cite{shin2010}
that different parts of the Fermi surface contributes differently 
towards to superconductivity.  These `non magnetic' compounds
are expected to have strong antiferromagnetic fluctuations, 
due to  Fermi surface nesting \cite{kontani}, affecting
the pairing and gap anisotropy.

Of the magnetic  superconductors,
HoNi$_{2}$B$_{2}$C, ErNi$_{2}$B$_{2}$C
and TmNi$_{2}$B$_{2}$C, show
considerable deviation of the gap from BCS behavior. 
Photoemission spectroscopy  
\cite{baba2008} on ErNi$_{2}$B$_{2}$C, 
point contact and Andreev reflection 
\cite{rybaltchencho1999,yanson2008} 
on HoNi$_{2}$B$_{2}$C and \cite{schultz2011} 
TmNi$_{2}$B$_{2}$C suggest  gap anisotropy on 
individual Fermi surface sheets, with magnitude variation
between $1.1-1.7 $meV
For  ErNi$_{2}$B$_{2}$C
and TmNi$_{2}$B$_{2}$C the deviations are visible even 
at the lowest temperature while in  HoNi$_{2}$B$_{2}$C, 
where $T_c > T_{AF}$, it is observed roughly above $T_{AF}$.
Existing measurements \cite{paul2000} 
suggest that DyNi$_{2}$B$_{2}$C, which has $T_{AF} > T_c$,
can be described in terms of `BCS' behavior (inconsistent with
what we suggest in Fig.12.(c)). We believe this
merits more careful probing.

\subsubsection{Thermal effects}

Since this paper is focused on the ground state we did not use
the full power of the Monte Carlo method. The detailed
finite temperature properties will be discussed separately,
here we provide a glimpse of the finite temperature phase
diagram that emerges. 
Beyond the `mean field' effect of the diminished
magnetic and superconducting order
at finite temperature one expects (i)~amplitude and phase
fluctuations of the pairing field to suppress the gap 
(at low $J$) with increasing $T$, and (ii)~the gap suppression
effect to be accelerated by the magnetic disorder which
would lead to strong spin flip scattering. 
These effects require a treatment well beyond mean field
theory and our Boltzmann sampling of thermal 
configurations $\{ \Delta_i, {\bf S}_i\}$
accomplishes that. The thermally generated disorder 
feeds back into the electrons to modify spectral
properties.

The superconducting $T_c$ falls quickly for $J > 0.5t$
and goes to zero at $J_c \sim t$, while the low $J$ gap
in the DOS closes at a scale $T_g$ that collapses at $J \sim 0.7t$.
We note that there cannot be a finite $T_{AF}$ in a 2D 
${\cal O}(3)$ invariant spin system, although the magnetic
correlation length grows exponentially as $T$ is lowered
below the indicated $T_{AF}$.
 
The data shown are at $U=4t$ where even 
calculations on $16 \times
16$ lattices are reliable. We are working on lower $U$,
which is physically more relevant, and will report the
thermal properties soon.

\section{Conclusions}

We have studied the interplay of superconductivity and magnetism 
in a two dimensional model involving an attractive Hubbard 
interaction and a Kondo like coupling to local moments. 
The ground state phase diagram is mapped out in terms of the 
attractive interaction, magnetic coupling, and electron filling. 
Over a range of magnetic coupling we observe a `gapless'
superconducting state existing generally with non-collinear 
magnetic order. For Neel order, superconductivity can 
coexist with magnetism but we do not observe a gapless phase
for the bandstructure we have chosen. 
We identify the origin of the gapless behavior in the 
participation of magnetic Bloch states in the pairing
process. The combination of pairing and magnetic `scattering'
leads to an effective `8 band' dispersion, with some bands
crossing the Fermi level at sufficiently large magnetic 
coupling. An approximate Green's function analysis
provides insight on these new bands.

The Monte Carlo technique used here in a limited way also
captures the thermal physics of the problem on fairly large lattices.
We have used it to map out the thermal phase diagram, determining the 
$T_{c}$ and $T_{AF}$ scales, presented here, 
and also the evolution of the spectral
features across the ordering transitions.  Results on this will be 
presented separately.  The approach here generalizes naturally 
to the problem of $d$-wave superconductivity
coexisting with magnetic order, as in some
heavy fermions and ferropnictides. We are
exploring these.

We acknowledge use of the High Performance Computing facility at HRI
and thank Nyayabanta Swain and Sauri Bhattacharyya for comments.

\section{Appendix: approximate dispersion in the gapless phase}

At small $\Delta_0/J$, in the gapless phase, we can write explicit
dispersions for the six bands that emerge from the Green's function
scheme. This involves solving for the three poles of the Green's
function, for each spin projection, at $\Delta_0 =0$, and then
calculating the small $\Delta_0$ corrections. In this spirit,
the three poles of the up spin Green's function at $\Delta_0=0$
are:
\begin{eqnarray}
 E_{1}^{0}({\bf k}) & = & E^{+}({\bf k}) \\ \nonumber 
 E_{2}^{0}({\bf k}) & = & E^{-}({\bf k}) \\ \nonumber
 E_{3}^{0}({\bf k}) & = & -\epsilon({\bf k})
\end{eqnarray}
where,
\begin{eqnarray}
 E^{\pm}({\bf k}) & = & { {(\epsilon_{\bf k}+\epsilon_{\bf k + Q}) \pm 
 \sqrt{(\epsilon_{\bf k} - \epsilon_{\bf k + Q})^{2} + 4 J^2S^2}} 
\over 2} \nonumber
 \\ 
\end{eqnarray}
The pole at $E^0_3({\bf k}) = -\epsilon({\bf k})$ has residue zero
(since it is artificial and cancels with a zero of the Green's function).
At $\Delta_0 \neq 0$, however, all the poles have non zero residues,
and the shifted poles are defined by $E_{\alpha}({\bf k})
= E^0_{\alpha}({\bf k}) + \eta_{\alpha}({\bf k})$, where:
\begin{eqnarray}
 \eta_{1}({\bf k}) & = & \frac{\Delta_{0}^{2}(E^{+}(\bf k) - \epsilon({\bf k+Q}))}
 {[(E^{+}({\bf k}) - E^{-}({\bf k}))(E^{+}({\bf k}) + \epsilon({\bf k}))-\Delta_{0}^{2}]} \nonumber 
\end{eqnarray}
\begin{eqnarray}
 \eta_{2}({\bf k}) & = & \frac{\Delta_{0}^{2}(E^{-}({\bf k})-\epsilon({\bf k+Q}))}
 {[(E^{-}({\bf k}) - E^{+}({\bf k}))(E^{-}({\bf k}) + \epsilon({\bf k}))-\Delta_{0}^{2}]} \nonumber
\end{eqnarray}
\begin{eqnarray}
 \eta_{3}({\bf k}) & = & \frac{-\Delta_{0}^{2}(\epsilon({\bf k}) + \epsilon({\bf k+Q}))}
 {[(\epsilon({\bf k}) + E^{-}({\bf k}))(\epsilon({\bf k}) + E^{+}({\bf k}))-\Delta_{0}^{2}]}
\end{eqnarray}

The above corrections correspond to the poles of $G_{\uparrow \uparrow}({\bf k}, i\omega_{n})$.
Similarly, corrections corresponding to the poles of $G_{\downarrow \downarrow}({\bf k}, i\omega_{n})$
can also be determined. Together with Eqn.(5) they give corrections to the six poles 
for the total $G({\bf k}, i\omega_{n})$.

%%%%%%%%%%%%%%%%%%%%%%%%%%%%%%%%%%%%%%%%%%%%%%%%%%%%

%--------------------------------------------------------------------

\end{document}